\documentclass[a4paper,11pt]{article}
\pdfoutput=1 % if your are submitting a pdflatex (i.e. if you have
\usepackage{jheppub} % for details on the use of the package, please
\usepackage[T1]{fontenc} % if needed

\usepackage{mathtools}
\usepackage{tensor}
\usepackage{braket}
\usepackage{cleveref}
\usepackage{booktabs}
\usepackage{diagbox}
\usepackage{transparent}
\usepackage{xcolor}

\crefname{table}{table}{tables}
\Crefname{table}{Table}{Tables}
\crefname{section}{section}{sections}
\Crefname{section}{Section}{Sections}
\crefname{figure}{figure}{figures}
\Crefname{figure}{Figure}{Figures}
\crefname{equation}{eq.}{eqs.}
\Crefname{equation}{Eq.}{Eqs.}
\crefname{appendix}{appendix}{appendices}
\Crefname{appendix}{Appendix}{Appendices}

\DeclareMathOperator{\tr}{Tr}
\DeclareMathOperator{\re}{\mathrm{Re}}
\DeclareMathOperator{\im}{\mathrm{Im}}
\newcommand{\vbar}{\big |}
\newcommand{\overbar}[1]{\mkern 1mu\overline{\mkern-1mu#1\mkern-1mu}\mkern 1mu}

\title{Supersymmetric geometry in non-supersymmetric effective field theory}

\author[a,b]{Nathaniel Craig,}
\author[a,c,d]{Andrew Fee}
\author[a]{and Yu-Tse Lee}

\affiliation[a]{Department of Physics, University of California, Santa Barbara, CA 93106, USA}
\affiliation[b]{Kavli Institute for Theoretical Physics, Santa Barbara, CA 93106, USA}
\affiliation[c]{Department of Physics, University of Chicago, Chicago, IL 60637, USA}
\affiliation[d]{Leinweber Institute for Theoretical Physics, University of Chicago, Chicago, IL 60637, USA}

\emailAdd{ncraig@ucsb.edu}
\emailAdd{feea@uchicago.edu}
\emailAdd{yutselee@ucsb.edu}

\abstract{We develop a geometric framework for non-supersymmetric effective gauge theories based on their nonlinear supersymmetrizations. We construct supersymmetric embeddings for most operators up to dimension six from constrained chiral and vector superfields, and formulate vector bundles using the superfields to systematically organize the operators under field redefinitions in the gauge sector. This formalism manifests a complex geometry underlying gauge operators and accommodates redefinitions across different spins.}

\begin{document}
\maketitle
\flushbottom

\section{Introduction}
\label{sec:intro}

Symmetry lays the foundation for established theories \cite{Einstein:1916vd, Wigner:1939cj, Yang:1954ek} and charts the frontier of active discoveries in physics \cite{Gaiotto:2014kfa}. The absence of symmetry in its most direct form does not hinder its utility, however. For instance, symmetries that are spontaneously broken give rise to universal phenomena like massless particles and soft theorems \cite{Nambu:1960tm, Goldstone:1961eq, Weinberg:1964ew, Weinberg:1965nx}, and are responsible for generating the masses of gauge bosons \cite{Englert:1964et, Higgs:1964pj, Guralnik:1964eu}. Going further, symmetries that are exact only when a theory is lifted by the extension of the field or coupling space provide powerful organizational principles in effective field theory (EFT) \cite{Gell-Mann:1960mvl, Sikivie:1980hm}, and serve as the basis behind naturalness and spurion analysis \cite{Oppenheimer:1956hfa, tHooft:1979rat, Seiberg:1993vc}. 

The central idea in the present paper is similar in spirit but more modest in ambition:
\begin{center}
    A supersymmetric lift systematically organizes operators in EFT.
\end{center}
As we will see, this is useful even for EFTs without supersymmetry (SUSY) at any scale. EFT operators suffer from a vast set of redundancies due to the fact that field redefinitions lead to physically equivalent theories \cite{Chisholm:1961tha, Kamefuchi:1961sb, Coleman:1969sm, Arzt:1993gz}. One way to resolve the redundancy is to categorize operators by derivative count and collect them into covariant tensors on the target space of the fields \cite{Meetz:1969as, Honerkamp:1971xtx, Honerkamp:1971sh, Ecker:1971xko, Alvarez-Gaume:1981exa, Alvarez-Gaume:1981exv, Boulware:1981ns, Howe:1986vm, Dixon:1989fj, Alonso:2015fsp, Alonso:2016oah}. This geometric classification of operators manifests invariance under a range of field reparameterizations and finds many conceptual and computational applications \cite{Alonso:2016btr, Helset:2018fgq, Nagai:2019tgi, Finn:2019aip, Helset:2020yio, Finn:2020nvn, Cohen:2020xca, Cohen:2021ucp, Alonso:2021rac, Cheung:2021yog, Cohen:2022uuw, Cheung:2022vnd, Alonso:2022ffe, Talbert:2022unj, Helset:2022tlf, Helset:2022pde, Craig:2023wni, Gattus:2023gep, Assi:2023zid, Craig:2023hhp, Alminawi:2023qtf, Jenkins:2023rtg, Jenkins:2023bls, Alonso:2023jsi, Cohen:2023ekv, Derda:2024jvo, Helset:2024vle, Gattus:2024ird, Cohen:2024bml, Lee:2024xqa, Li:2024ciy, Cohen:2024fak, Aigner:2025xyt, Cohen:2025dex, Assi:2025fsm, Craig:2025uoc, Cohen:2025prs, Alminawi:2025pwg, Alonso:2025ksv, Delgado:2026zpz}. Nevertheless, lingering limitations on spins and derivatives in the allowed redefinitions motivate the introduction of SUSY as a further organizing principle, as the structure of superspace can illuminate relations among a wider class of operators based on their supersymmetric embeddings.\footnote{Supersymmetric lifts of non-supersymmetric EFTs have also been used to good effect in understanding aspects of operator non-renormalization \cite{Elias-Miro:2014eia}.} 

In \cite{Lee:2024xqa}, generic scalars and fermions are nonlinearly supersymmetrized by filling in their would-be superpartners with spectator Goldstino states. Restricting to non-derivative couplings, the corresponding constrained superfields \cite{Volkov:1973ix, Ivanov:1978mx, Rocek:1978nb, Lindstrom:1979kq, Casalbuoni:1988xh, Komargodski:2009rz, Kuzenko:2010ef} define a geometry that accounts for derivative redefinitions of the embedded particles, due to the fact that the superfield target space is Kähler. Importantly, SUSY is not a requirement on the original theory; the newly introduced Goldstino couplings are suppressed by the scale of SUSY breaking and can be made arbitrarily small. Rather, that the theory \emph{can be} supersymmetrized in this fashion suffices to reveal an underlying Kähler geometry, even if not immediately apparent from component fields. However, demanding that no derivative superfield couplings are present places restrictions on the specific forms of some scalar and fermionic operators in the original theory; these are the conditions for the theory to be approximated by a nonlinear sigma model of constrained chiral superfields.

In the present work, we extend this observation to a more general setting. First, we expand the range of EFTs supersymmetrized, including gauge theories, by introducing constrained vector superfields \cite{Komargodski:2009rz, DallAgata:2016syy} and turning on derivative couplings to the SUSY-breaking Goldstino superfield. The auxiliary field contained within remains essentially as a non-propagating field in the construction. We explicitly treat all operators up to dimension six; from Grassmann counting on superspace, supersymmetrization will generally become easier at higher dimensions. This demonstrates that given a generic EFT, one can find a nonlinearly supersymmetric version that is arbitrarily close so that deviations from the original are minimal. Some requirements on the scalar sector remain, but they are an artifact of this particular construction and do not define the limits of what is supersymmetrizable.\footnote{The related task of operator counting and construction in linear SUSY is undertaken in \cite{Delgado:2022bho, Delgado:2023ivp, Delgado:2023ogc, Delgado:2024ivu, Delgado:2025oev}.}

Then, we leverage the superfield approximation of the EFT to elucidate its latent geometry, with attention to gauge operators. Although restrictions on the theory have been relaxed compared to \cite{Lee:2024xqa}, its target space retains notable structures that extend existing constructions, particularly in the gauge sector \cite{Helset:2018fgq, Helset:2022tlf, Helset:2022pde, Assi:2023zid, Derda:2024jvo, Assi:2025fsm}. Using the gauge kinetic function, we explicitly incorporate CP-odd operators together with the even ones to form a complex geometry that captures redefinitions involving the field strength. Moreover, we derive an underlying gauge boson geometry that enables redefinitions across different spins, but nevertheless disentangles them so that the field strength emerges. The derivative couplings introduced for supersymmetrization are put into geometric form by devising appropriate covariant derivatives. This organization of gauge operators under field redefinitions leverages the properties of superspace, and would not have been directly obvious in component form.

The rest of the paper is organized as follows. \Cref{sec:susy} lays out a detailed procedure for supersymmetrizing a generic EFT by assembling operators out of constrained chiral and vector superfields. \Cref{sec:geometry} uses both the field strength and vector superfields to formulate a hierarchy of vector bundles in the gauge sector, discussing their relation and variations. \Cref{sec:conc} summarizes the results and suggests directions for future work. The Lie algebra and spinor conventions adopted can be found in \cref{app:conv}.

\section{Supersymmetrizing an effective field theory}
\label{sec:susy}

There is no unique approximation of an EFT by a supersymmetric one, although the SUSY must be nonlinear for a generic EFT. Our strategy is to embed each degree of freedom in the original theory into a constrained superfield, together with a multi-particle superpartner state containing the Goldstino and suppressed by the scale $\sqrt{f}$ of SUSY breaking. We require that $\sqrt{f}$ parametrically exceeds the EFT cutoff $E$. The superfields are next combined so that the desired operator appears, possibly dressed with an auxiliary field, in an appropriate $D$- or $F$-term on superspace. The auxiliary fields are then integrated out to leave behind a supersymmetric theory that reproduces the EFT at leading order in $f$ and relegates Goldstino couplings to higher orders.

\subsection{From fields to constrained superfields}

We consider a gauge group with Lie algebra $\mathfrak{g}$, generators $T_A = (T_A)^\dagger$ and structure constants $\tensor{f}{_{AB}^C}$. Aside from the $\mathfrak{g}$-valued gauge bosons $A_\mu = A_\mu^B T_B$, the EFT can contain real or complex scalars $\phi^a$ and two-component fermions $\psi^p$, each a multiplet in a representation $R^a$ or $R^p$.\footnote{The Lie algebra index $B$ may subsequently be suppressed to reduce clutter. We use the symbol $\dagger$ for the conjugate transpose of objects from multi-dimensional representations.}

Meanwhile, the supersymmetric theory must contain at least one additional particle, namely the Goldstino $G$. It will later be arranged to decouple from the other particles at leading order in $f$, so that the theory approximates the original. At the moment, we focus on making sure SUSY is actually realized, a task best suited for superspace. Hence, we seek an equivalent parameterization of the theory with superfields, whose components are to be filled in with the various particles and nothing more, besides possibly some auxiliary fields. This can be achieved using the technology of constrained superfields.

The detailed SUSY breaking mechanism at the scale $\sqrt{f}$ is largely unimportant at energies below $E$. For simplicity, we assume that an auxiliary field $F$ acquires a non-zero vacuum expectation value (vev), $\langle F \rangle \sim - f$. Then this auxiliary field, together with the Goldstino, can be assembled into a chiral superfield
\begin{equation}
    X = \frac{G^2}{2F} + \sqrt{2} \theta G + \theta^2 F \, , \label{eq:X}
\end{equation}
where component fields on the right hand side are functions of $y^\mu = x^\mu + i \theta \sigma^\mu \overbar{\theta}$. With no independent sgoldstino available, the SUSY transformation of $X$ fixes its lowest component essentially as above and results in the nilpotent identity
\begin{equation}
    X^2 = 0 \, .
\end{equation}
Conversely, a chiral superfield $X$ constrained by $X^2 = 0$ has \cref{eq:X} as the general solution, provided its auxiliary field obtains a vev.

Similarly, a complex scalar $\phi^a$ and a fermion $\psi^p$ can be packaged together with auxiliary fields $F^a$ and $F^p$ into chiral superfields
\begin{align}
    Z^a &= \phi^a + \sqrt{2} \theta \left [ \frac{\partial_\mu \phi^a}{\overbar{F}} \, i \sigma^\mu \overbar{G} + \frac{F^a}{F} \, G + \mathcal{O} \left ( \frac{1}{f^2} \right ) \right ] + \theta^2 F^a \, , \label{eq:Z} \\
    Y^p &= \left [ \frac{\psi^p G}{F} + \mathcal{O} \left ( \frac{1}{f^2} \right ) \right ] + \sqrt{2} \theta \psi^p + \theta^2 F^p \, . \label{eq:Y}
\end{align}
In counting powers of $f$, we include factors of $F$ and $\overbar{F}$ which acquire vevs. The superpartners, whose leading terms in $f$ are shown, are given by precise combinations of the other component fields with $G$ and $F$ from $X$ such that the constraints
\begin{align}
    \overbar{X} X D_\alpha Z^a &= 0 \, , \label{eq:Z_constraint} \\
    X Y^p &= 0 \, ,
\end{align}
are satisfied, where $D_\alpha = \partial_\alpha + i \sigma^\mu_{\alpha \dot{\alpha}} \overbar{\theta}^{\dot{\alpha}} \partial_\mu$ is the chiral covariant derivative. While inessential, we have opted to impose \cref{eq:Z_constraint} on $Z^a$ \cite{DallAgata:2015zxp} instead of the stronger constraint $\overbar{X} D_\alpha Z^a = 0$ \cite{Komargodski:2009rz}, so that the auxiliary field $F^a$ survives and lends more flexibility in generating the scalar potential later. We will later also arrange for consistency that $F^a$ and $F^p$ do not acquire vevs, so that the Goldstino indeed sits entirely in $X$.

For a real scalar $\varphi^a$, we should further enforce $\overbar{X} X \im Z^a = 0$ in addition to \cref{eq:Z_constraint}. This again yields \cref{eq:Z}, but now with $\phi^a = \varphi^a + \mathcal{O}(f^{-2})$ so that the imaginary scalar is further eliminated \cite{DallAgata:2016syy}. We will henceforth stick to complex scalars in the EFT, knowing that the same computations for real scalars amount to a simple replacement.

The gauge bosons $A_\mu$, together with the auxiliary field $D$, make up a vector superfield
\begin{align}
    V &= \frac{\theta \sigma^\mu \overbar{G}}{\sqrt{2} \overbar{F}} \, A_\mu - \frac{\overbar{\theta} \overbar{\sigma}^\mu G}{\sqrt{2} F} \, A_\mu + \theta \sigma^\mu \overbar{\theta} A_\mu + \theta^2 \overbar{\theta} \left [ \lambda^\dagger + \overbar{\sigma}^\mu \partial_\mu \left ( \frac{i \sigma^\nu \overbar{G}}{2 \sqrt{2} \overbar{F}} \, A_\nu \right ) \right ] \notag \\
    &\quad + \overbar{\theta}^2 \theta \left [ \lambda - \sigma^\mu \partial_\mu \left ( \frac{i \overbar{\sigma}^\nu G}{2 \sqrt{2} F} \, A_\nu \right ) \right ] + \frac{1}{2} \theta^2 \overbar{\theta}^2 D + \mathcal{O} \left ( \frac{1}{f^2} \right ) \, , \label{eq:V}
\end{align}
where the component fields are functions of $x^\mu$. What would have been the gaugino is now a Goldstino state
\begin{equation}
    \lambda_\alpha = \frac{D}{\sqrt{2} F} \, G_\alpha + \frac{F_{\mu\nu}}{\sqrt{2} F} \, (\sigma^{\mu\nu} G)_\alpha + \mathcal{O} \left ( \frac{1}{f^3} \right ) \, , \label{eq:lambda}
\end{equation}
where $F_{\mu\nu} = \partial_\mu A_\nu - \partial_\nu A_\mu - i [A_\mu, A_\nu]$ is the field strength. The vector superfield $V$ yields the chiral field strength superfield
\begin{align}
    W_\alpha &= - \frac{1}{8} \, \overbar{D}_{\dot{\alpha}} \, \overbar{D}^{\dot{\alpha}} \left ( e^{-2V} D_\alpha \, e^{2V} \right ) \\
    &= \lambda_\alpha(y) + \theta_\alpha D + (\sigma^{\mu\nu} \theta)_\alpha F_{\mu\nu} - \theta^2 (i \sigma^\mu \nabla_\mu \lambda^\dagger)_\alpha \, . \label{eq:W}
\end{align}
where $\nabla_\mu = \partial_\mu - i A_\mu \, \cdot$ is the gauge covariant derivative and $\cdot$ denotes the group action.\footnote{With $\lambda_\alpha$ in the adjoint representation, $A_\mu \cdot \lambda_\alpha = [A_\mu, \lambda_\alpha]$.} \Cref{eq:V,eq:lambda} result from imposing the respective constraints
\begin{align}
    X V &= 0 \, , \label{eq:V_constraint} \\
    X W_\alpha &= 0 \, .
\end{align}
\Cref{eq:V_constraint} is a partial gauge choice akin to Wess-Zumino gauge and restricts gauge transformations
\begin{equation}
    e^{2V} \rightarrow e^{-2i \Omega^\dagger} e^{2V} e^{2i \Omega} \, ,
\end{equation}
to chiral superfields $\Omega$ that satisfy $X \im \Omega = 0$. Then $\Omega = \omega + \mathcal{O}(f^{-1})$ contains only an independent real scalar $\omega$ that enacts component gauge transformations on $A_\mu$. This choice serves to preserve the constraint \cref{eq:Z_constraint} under the transformation
\begin{equation}
    Z^a \rightarrow e^{-2i \Omega} \cdot Z^a = e^{-2i \Omega^A T_A^{\,a}} Z^a \, ,
\end{equation}
where $T_A^{\,a}$ are the generators of $R^a$ \cite{Komargodski:2009rz}.

Each field in the EFT has now been embedded as a component in a corresponding matter superfield. However, additional fields have been introduced along the way, namely the Goldstino $G$ and auxiliary field $F$ from the SUSY breaking sector, and other auxiliary fields $F^a$, $F^p$ and $D^A$. We now proceed to construct supersymmetric versions of various EFT operators using the superfields, keeping track of contributions from the additional fields.

\subsection{From operators to super-operators}

Our present task is to combine constrained superfields so that a gauge invariant EFT operator $O_i$ appears as a $D$- or $F$-term, perhaps along with other operators that are necessary for SUSY. While this is possible for some $O_i$, it becomes considerably easier if we further allow a factor of $\theta^2 F$ coming from the Goldstino superfield $X$. To be concrete, we will engineer super-operators $X S_i$\footnote{Or their Hermitian conjugates.} with suitable matter superfields in the gauge singlet $S_i$ such that
\begin{equation}
    \frac{1}{f} \, X S_i \, \vbar_{\theta^2 \overbar{\theta}^2} = \frac{1}{f} \, F O_i + \mathcal{O} \left ( \frac{1}{f} \right ) \, .
\end{equation}
Previewing what is to come, when we eventually integrate out $F$ from the theory\footnote{We use the notation $\overbar{G} \overbar{\sigma}^\mu \overset{\text{\tiny$\leftrightarrow$}}{\partial}_\mu G = \overbar{G} \overbar{\sigma}^\mu (\partial_\mu G) - (\partial_\mu \overbar{G}) \overbar{\sigma}^\mu G$.}
\begin{align}
    \mathcal{L} \supset \mathcal{L}_X &= \int d^4 \theta \; \overbar{X} X + \left [ \int d^2 \theta \; f X - \int d^4 \theta \, \sum_i \frac{c_i}{f} \, X S_i + \text{c.c.} \right ] \\
    &= \overbar{F} F + \left [ \left ( f - \frac{1}{f} \sum_i c_i O_i \right ) F + \text{c.c.} \right ] - \frac{i}{2} \, \overbar{G} \overbar{\sigma}^\mu \overset{\text{\tiny$\leftrightarrow$}}{\partial}_\mu G + \mathcal{O} \left ( \frac{1}{f} \right ) \, ,
\end{align}
we will obtain $c_i O_i$ at order $f^0$ after due care with the other terms. Another advantage of this protocol is that all additional operators in $X S_i$ due to SUSY are Goldstino terms sub-leading in $f$, so that there is a clean supersymmetrization of individual EFT operators at leading order, as opposed to e.g. the two classes of fermion kinetic terms in \cite{Lee:2024xqa} which share the same super-operator when $X$ is not thus utilized.

Since $X$ supplies $\theta^2$, we need two factors of $\overbar{\theta}$ in $S_i$. They can come from the chiral superfields $Z^a$, $Y^p$, $W_\alpha$, or e.g. their gauge and chiral covariant derivative by
\begin{equation}
    \nabla_\alpha = D_\alpha + (e^{-2V} D_\alpha \, e^{2V}) \, \cdot \, . \label{eq:gauge_chiral_deriv}
\end{equation}
Options for obtaining scalars in $O_i$ include
\begin{align}
    Z^a \, \vbar_{f^0} &= \phi^a(x) + \theta^2 F^a + \mathcal{O}(\theta \overbar{\theta}) \, , \\
    \nabla_\alpha Z^a \, \vbar_{f^0} &= 2 \theta_\alpha F^a + 2 (i \sigma^\mu \overbar{\theta})_\alpha \nabla_\mu \phi^a + \mathcal{O}(\theta \overbar{\theta}) \, ,
\end{align}
where the components are functions of $x^\mu$, and $\mathcal{O}(\theta \overbar{\theta})$ indicates that the remaining terms at order $f^0$ contain at least one $\theta$ and one $\overbar{\theta}$. Any other term in $Z^a$ or $\nabla_\alpha Z^a$ is $\mathcal{O}(f^{-1})$. Likewise, for fermions and gauge bosons, we have respectively
\begin{align}
    Y^p \, \vbar_{f^0} &= \sqrt{2} \theta \psi^p + \theta^2 F^p + \mathcal{O}(\theta^2 \overbar{\theta}) \, , \\
    \nabla_\alpha Y^p \, \vbar_{f^0} &= \sqrt{2} \psi^p_\alpha + 2 \theta_\alpha F^p + \mathcal{O}(\theta \overbar{\theta}) \, ,
\end{align}
and
\begin{align}
    W_\alpha \, \vbar_{f^0} &= \theta_\alpha D + (\sigma^{\mu\nu} \theta)_\alpha F_{\mu\nu} + \mathcal{O}(\theta^2 \overbar{\theta}) \, , \\
    \nabla_\beta W_\alpha \, \vbar_{f^0} &= - \, \epsilon_{\beta \alpha} D + (\sigma^{\mu\nu})_\beta^{\;\;\gamma} \epsilon_{\gamma \alpha} F_{\mu\nu} + \mathcal{O}(\theta \overbar{\theta}) \, .
\end{align}
We see that the $\overbar{\theta}$ components in $\nabla_\alpha Z^a$, $Y^{\overbar{p} \dagger}$ and $W^\dagger_{\dot{\alpha}}$ are useful for this purpose. We will elect not to use the second order derivative $\nabla^\alpha \nabla_\alpha$; it turns out that other covariant derivatives on the superfields, namely $\overbar{D}_{\dot{\alpha}}$ and
\begin{equation}
    \nabla_\mu = \partial_\mu + \frac{i}{4} \, \overbar{D}_{\dot{\alpha}} \left ( \overbar{\sigma}_\mu^{\dot{\alpha} \alpha} e^{-2V} D_\alpha \, e^{2V} \right ) \, \cdot = \partial_\mu - i A_\mu \cdot + \, \mathcal{O}(\theta \overbar{\theta}) \, ,
\end{equation}
are also not needed up to dimension six in the EFT.

We can now form the super-operators
\begin{align}
    X \nabla^\alpha Z^a \nabla_\alpha Z^b &= \theta^2 \overbar{\theta}^2 \, F \, \Big [ 4 \, \nabla^\mu \phi^a \nabla_\mu \phi^b \Big ] + \mathcal{O}(f^0) \, , \\
    \overbar{X} Y^p Y^q &= \theta^2 \overbar{\theta}^2 \, \overbar{F} \, \Big [ - (\psi^p \psi^q) \Big ] + \mathcal{O}(f^0) \, , \\
    X Y^{\overbar{p} \dagger} \nabla^\alpha Y^q \nabla_\alpha Z^a &= \theta^2 \overbar{\theta}^2 \, F \, \Big [ 2i \, (\psi^{\overbar{p} \dagger} \overbar{\sigma}^\mu \psi^q) \nabla_\mu \phi^a \Big ] + \mathcal{O}(f^0) \, , \\
    \overbar{X} \Big [ 2 \, W^{[A|\alpha} W^{|B|\beta} \nabla_\alpha^{\vphantom{C]}} W^{|C]}_\beta \hspace{3.9em} & \notag \\
    - \, W^{[A|\alpha} W^{|B|}_\alpha \nabla^\beta W^{|C]}_\beta \Big ] &= \theta^2 \overbar{\theta}^2 \, \overbar{F} \, \Big [ - 4 i \, F^{\mu [A|}_{\;\;\nu} F^{\nu\rho |B|} F_{\rho\mu}^{+ |C]} \Big ] + \mathcal{O}(f^0) \, , \label{eq:3F} \\
    \overbar{X} Y^p \Big [ 2 \, (\nabla^\alpha Y^q) W_\alpha^A + Y^q (\nabla^\alpha W_\alpha^A) \Big ] &= \theta^2 \overbar{\theta}^2 \, \overbar{F} \, \Big [ 2 \, (\psi^p \sigma^{\mu\nu} \psi^q) F_{\mu\nu}^A \Big ] + \mathcal{O}(f^0) \, ,
\end{align}
so that each $X S_i$ yields a single $F O_i$ at leading order after numerical rescaling.\footnote{The Lie algebra indices in \cref{eq:3F} are anti-symmetrized to eliminate leading terms containing the auxiliary field $D^A$.} Here, $F_{\mu\nu}^\pm = (F_{\mu\nu} \mp i \widetilde{F}_{\mu\nu}) / 2$ contains the dual field strength $\widetilde{F}_{\mu\nu} = \epsilon_{\mu\nu\rho\sigma} F^{\rho\sigma} / 2$. We can also dress $X S_i$ with further factors of e.g. $Z^a$ and $\nabla_\alpha Y^p$, so that corresponding factors of $\phi^a$ and $\psi^p_\alpha$ are inserted into $F O_i$. While these are sufficient to cover a broad range of operators, there is a notable exception below for which we will use
\begin{equation}
    \overbar{X} W^{A\alpha} W_\alpha^B = \theta^2 \overbar{\theta}^2 \, \overbar{F} \, \Big [ - F^{\mu\nu A} F_{\mu\nu}^{+ B} + D^A D^B \Big ] + \mathcal{O}(f^0) \, , \\
\end{equation}
at the cost of generating other operators upon integrating out $D^A$.

Whichever $S_i$ is ultimately utilized, the various representations that it contains are to be combined as in $O_i$ for gauge invariance. We have generally left the combinations unspecified; they should be tailored to the specific EFT of interest. For example, if the EFT contains a charged fermion $\psi^r$ and a real scalar $\varphi^c$, we have
\begin{equation}
    X \left ( Y^{\overbar{r} \dagger} e^{2V} \nabla^\alpha Y^r \right ) \nabla_\alpha Z^c = \theta^2 \overbar{\theta}^2 \, F \, \Big [ 2i \, (\psi^{\overbar{r} \dagger} \overbar{\sigma}^\mu \psi^r) \nabla_\mu \varphi^c \Big ] + \mathcal{O}(f^0) \, , \label{eq:super_operator_example}
\end{equation}
as the explicitly gauge invariant supersymmetrization of a fermion kinetic term. The insertion of $e^{2V} = [1 + \mathcal{O}(\theta \overbar{\theta})] + \mathcal{O}(f^{-1})$ in $S_i$ does not affect the embedded $O_i$ by construction.

The dimension five and six EFT operators that can be straightforwardly supersymmetrized under this scheme are:\footnote{In the $\boxed{\text{operator classes}}$, $\phi^n$ refers to an unspecified number of scalars and $\mathcal{X}$ refers to the field strength.}
\begin{itemize}
    \item[$\boxed{\psi^4}$] $(\psi^p \psi^r) (\psi^q \psi^s)$ and $(\psi^{\overbar{p} \dagger} \psi^{\overbar{r} \dagger}) (\psi^q \psi^s)$ \\ are covered by $\overbar{X} Y^q Y^s$ dressed with $\nabla^\alpha Y^p \nabla_\alpha Y^r$ and $\nabla_{\dot{\alpha}} Y^{\overbar{p} \dagger} \nabla^{\dot{\alpha}} Y^{\overbar{r} \dagger}$.
    
    \item[$\boxed{\psi^2 \phi^n}$] $(\psi^p \psi^q) \phi^a \phi^b$, $(\psi^p \psi^q) \phi^{\overbar{a} \dagger} \phi^b$, $(\psi^p \psi^q) \phi^a \phi^b \phi^c$, $\ldots$ with any combination of scalars \\ are covered by $\overbar{X} Y^p Y^q$ dressed with $Z^a Z^b$, $Z^{\overbar{a} \dagger} Z^b$, $Z^a Z^b Z^c$ and so on.
    
    \item[$\boxed{\psi^2 \phi^n \nabla}$] $(\psi^{\overbar{p} \dagger} \overbar{\sigma}^\mu \psi^q) \nabla_\mu \phi^a$, $(\psi^{\overbar{p} \dagger} \overbar{\sigma}^\mu \psi^q) (\nabla_\mu \phi^a) \phi^b$ and $(\psi^{\overbar{p} \dagger} \overbar{\sigma}^\mu \psi^q) (\nabla_\mu \phi^a) \phi^{\overbar{b} \dagger}$ \\ are covered by $X Y^{\overbar{p} \dagger} \nabla^\alpha Y^q \nabla_\alpha Z^a$ possibly dressed with $Z^b$ or $Z^{\overbar{b} \dagger}$.
    
    \item[$\boxed{\psi^2 \mathcal{X} \phi^n}$] $(\psi^p \sigma^{\mu\nu} \psi^q) F_{\mu\nu}$, $(\psi^p \sigma^{\mu\nu} \psi^q) F_{\mu\nu} \phi^a$ and $(\psi^p \sigma^{\mu\nu} \psi^q) F_{\mu\nu} \phi^{\overbar{a} \dagger}$ \\ are covered by $\overbar{X} Y^p [ 2 (\nabla^\alpha Y^q) W_\alpha + Y^q (\nabla^\alpha W_\alpha) ]$ possibly dressed with $Z^a$ or $Z^{\overbar{a} \dagger}$.
    
    \item[$\boxed{\phi^n \nabla^2}$] $(\nabla^\mu \phi^a \nabla_\mu \phi^b) \phi^c$, $(\nabla^\mu \phi^a \nabla_\mu \phi^b) \phi^{\overbar{c} \dagger}$, $(\nabla^\mu \phi^a \nabla_\mu \phi^b) \phi^{\overbar{c} \dagger} \phi^d$, $\ldots$ containing $\nabla^\mu \phi^a \nabla_\mu \phi^b$ \\ are covered by $X \nabla^\alpha Z^a \nabla_\alpha Z^b$ dressed with factors of $Z$ or $Z^\dagger$.\footnote{These are useful for avoiding any integrability constraint analogous to \cref{eq:scalar_kinetic_condition} when supersymmetrizing the kinetic terms of real scalars $\varphi^a$.}
    
    \item[$\boxed{\mathcal{X}^2 \phi^n}$] $F^{\mu\nu} F_{\mu\nu}^\pm \phi^a$, $F^{\mu\nu} F_{\mu\nu}^\pm \phi^a \phi^b$ and $F^{\mu\nu} F_{\mu\nu}^\pm \phi^{\overbar{a} \dagger} \phi^b$ are covered by $\overbar{X} W^\alpha W_\alpha$ dressed with \\ factors of $Z$ or $Z^\dagger$, but incur additional $\overbar{F} D^A D^B$ terms.
    
    \item[$\boxed{\mathcal{X}^3}$] $\tr F^\mu_{\;\;\nu} F^{\nu\rho} F_{\rho\mu}^\pm$ is covered by $\overbar{X} \tr \, [ 2 W^\alpha W^\beta \nabla_\alpha W_\beta - W^\alpha W_\alpha \nabla^\beta W_\beta ]$.
\end{itemize}

Missing from this list are
\begin{align}
    (\psi^{\overbar{p} \dagger} \overbar{\sigma}^\mu \nabla_\mu \psi^q) ( \leq 2 \text{ scalars} ) &\in \boxed{\psi^2 \phi^n \nabla} \; , \label{eq:fermion_kinetic} \\
    ( \leq 6 \text{ scalars} ) &\in \boxed{\phi^n} \; , \label{eq:scalar_potential} \\
    (\nabla^\mu \phi^{\overbar{a} \dagger} \nabla_\mu \phi^b) ( \leq 2 \text{ scalars} ) &\in \boxed{\phi^n \nabla^2} \; , \label{eq:scalar_kinetic}
\end{align}
which are not easy to embed cleanly in a $D$- or $F$-term, even if a factor of $\theta^2$ comes from $X$. For these operators, we will not invoke the auxiliary field $F$ in $X$, or derivatives of the matter superfields $Z^a$, $Y^p$ and $W_\alpha$. The non-derivative supersymmetric Lagrangian\footnote{This is roughly the gauged version of the theory in \cite{Lee:2024xqa}.}
\begin{align}
    \mathcal{L}_{\text{non-}X} &= \int d^4 \theta \; \bigg [ \, g(Z^\dagger, e^{2V} Z) + Y^{\overbar{p} \dagger} \, k_{\overbar{p}q}(Z^\dagger, e^{2V} Z) \, [e^{2V} Y^q] \\
    &\hspace{14.45em} + Y^{\overbar{p} \dagger} \, Y^{\overbar{r} \dagger} \, r_{\overbar{p} q \overbar{r} s}(Z^\dagger, e^{2V} Z) \, [e^{2V} Y^q] \, [e^{2V} Y^s] \, \bigg ] \notag \\
    &\quad + \left \{ \int d^2 \theta \; \Big [ \, w(Z) + f_p(Z) \, Y^p + m_{pr}(Z) \, Y^p Y^r + h_{AB}(Z) \, W^{\alpha A} W_\alpha^B \Big ] + \text{c.c.} \right \} \subset \mathcal{L} \, , \notag
\end{align}
where $Z$-dependent functions should transform appropriately for gauge invariance, has the component form\footnote{Commas in the subscript denote partial derivatives with respect to fields. We may leave the comma in the scalar kinetic term implicit, like in \cref{eq:scalar_kinetic_condition}.}
\begin{align}
    \mathcal{L}_{\text{non-}X} &= g_{,\overbar{a}b} \, \nabla_\mu \phi^{\overbar{a} \dagger} \, \nabla^\mu \phi^b - \frac{i}{2} \, k_{\overbar{p}q} \, (\psi^{\overbar{p} \dagger} \overbar{\sigma}^\mu \overset{\text{\tiny$\leftrightarrow$}}{\nabla}_\mu \psi^q) + r_{\overbar{p} q \overbar{r} s} \, (\psi^{\overbar{p} \dagger} \psi^{\overbar{r} \dagger}) (\psi^q \psi^s) \notag \\
    &\quad + g_{,\overbar{a}b} \, F^{\overbar{a} \dagger} F^b + k_{\overbar{p}q} \, F^{\overbar{p} \dagger} F^q + g_{,a} \, D \cdot \phi^a \notag \\
    &\quad + \bigg \{ - \frac{i}{2} \, k_{\overbar{p}q,a} \, (\psi^{\overbar{p} \dagger} \overbar{\sigma}^\mu \psi^q) \, \nabla_\mu \phi^a - m_{pr} \, (\psi^p \psi^r) - h_{AB} \, F^{\mu\nu A} F_{\mu\nu}^{+B} \notag \\
    &\hspace{3em} + w_{,a} \, F^a + f_p \, F^p + h_{AB} \, D^A D^B + \text{c.c.} \bigg \} + \mathcal{O} \left ( \frac{1}{f} \right ) \, ,
\end{align}
where the same functions are now $\phi$-dependent; without $F$ from $X$, all Goldstino couplings are $\mathcal{O}(f^{-1})$. This essentially generates \cref{eq:fermion_kinetic} with no restrictions, since among the three derivative structures
\begin{equation}
    (\psi^{\overbar{p} \dagger} \overbar{\sigma}^\mu \psi^q) \nabla_\mu \, , \enspace \psi^{\overbar{p} \dagger} \overbar{\sigma}^\mu (\nabla_\mu \psi^q) \, , \enspace (\nabla_\mu \psi^{\overbar{p} \dagger}) \overbar{\sigma}^\mu \psi^q \enspace \text{in} \enspace \boxed{\psi^2 \phi^n \nabla} \, ,
\end{equation}
only two are non-redundant. The scalar kinetic term \cref{eq:scalar_kinetic} is also produced, but with a metric\footnote{Adapting notation from the next section, $g_{\overbar{a}b}$ really refers to the block in the matrix $g_{\overbar{I}J}$ where $I \in I_a$ and $J \in I_b$.}
\begin{equation}
    g_{\overbar{a}b} (\phi^\dagger, \phi) \equiv g_{,\overbar{a}b} (\phi^\dagger, \phi) = \frac{\partial}{\partial \phi^{\overbar{a} \dagger} \vphantom{\phi^b}} \frac{\partial}{\partial \phi^b \vphantom{\phi^{\overbar{a} \dagger}}} \, g(\phi^\dagger, \phi) \, , \label{eq:scalar_kinetic_condition}
\end{equation}
that must be Kähler. Similarly, a restricted form of the scalar potential \cref{eq:scalar_potential} will arise after the auxiliary fields $F^q$, $F^b$ and $D^A$ have been integrated out. Note that $\mathcal{L}_{\text{non-}X}$ also accounts for all renormalizable operators, and provides an alternative supersymmetrization of four-fermion and holomorphic operators like
\begin{align}
    (\psi^{\overbar{p} \dagger} \psi^{\overbar{r} \dagger}) (\psi^q \psi^s) &\in \boxed{\psi^4} \, , \\
    (\psi^p \psi^q) \phi^a \phi^b \, , \enspace (\psi^p \psi^q) \phi^a \phi^b \phi^c &\in \boxed{\psi^2 \phi^n} \, , \\
    F^{\mu\nu} F_{\mu\nu}^+ \phi^a \, , \enspace F^{\mu\nu} F_{\mu\nu}^+ \phi^a \phi^b &\in \boxed{\mathcal{X}^2 \phi^n} \, ,
\end{align}
without invoking $X$ and possibly incurring additional $F D^A D^B$ terms for operators on the last line.

We have now constructed a supersymmetric embedding for each EFT operator at dimension six and below. The use of the Goldstino superfield $X \supset \theta^2 F$ allows us to mostly achieve a clean supersymmetrization, with the scalar kinetic term and potential as notable exceptions. We now proceed to integrate out all auxiliary fields and derive the final form of the nonlinearly supersymmetrized theory, with attention to how the scalar potential and the operators $O_i$ embedded in $S_i$ emerge.

\subsection{The physical supersymmetric theory}

Gathering all pieces, the theory $\mathcal{L} = \mathcal{L}_X + \mathcal{L}_{\text{non-}X}$ of constrained superfields reads
\begin{align}
    \mathcal{L} &= g_{\overbar{a}b} \, \nabla_\mu \phi^{\overbar{a} \dagger} \, \nabla^\mu \phi^b - \Big [ \, h_{AB} \, F^{\mu\nu A} F_{\mu\nu}^{+B} + \text{c.c.} \, \Big ] \notag \\
    &\quad - \frac{i}{2} \, k_{\overbar{p}q} \, (\psi^{\overbar{p} \dagger} \overbar{\sigma}^\mu \overset{\text{\tiny$\leftrightarrow$}}{\nabla}_\mu \psi^q) - \Big [ \, \frac{i}{2} \, k_{\overbar{p}q,a} \, (\psi^{\overbar{p} \dagger} \overbar{\sigma}^\mu \psi^q) \, \nabla_\mu \phi^a + \text{c.c.} \, \Big ] \notag \\
    &\quad - \Big [ \, m_{pr} \, (\psi^p \psi^r) + \text{c.c.} \, \Big ] + r_{\overbar{p} q \overbar{r} s} \, (\psi^{\overbar{p} \dagger} \psi^{\overbar{r} \dagger}) (\psi^q \psi^s) + \mathcal{L}_{\text{aux}+G} \label{eq:direct_SUSY_operators} \\
    \mathcal{L}_{\text{aux}+G} &= \overbar{F} F + g_{\overbar{a}b} \, F^{\overbar{a} \dagger} F^b + k_{\overbar{p}q} \, F^{\overbar{p} \dagger} F^q + \Big [ \Big ( h_{AB} + \frac{1}{f} \, \zeta_{AB}(c_{\text{nh}}) \, F \Big ) \, D^A D^B + \text{c.c.} \, \Big ] \notag \\
    &\quad + \Big [ \Big ( f - \frac{1}{f} \sum_i c_i O_i \Big ) \, F + w_{,a} \, F^a + f_p \, F^p + \text{c.c.} \, \Big ] - p_A \, D^A \notag \\
    &\quad - \frac{i}{2} \, \overbar{G} \overbar{\sigma}^\mu \overset{\text{\tiny$\leftrightarrow$}}{\partial}_\mu G + \mathcal{O} \left ( \frac{1}{f} \right ) \, ,
\end{align}
in component form. Other than $f$ and the Wilson coefficients $c_i$, the lowercase symbols denote $\phi$-dependent functions that should transform appropriately for gauge invariance. From top to bottom:
\begin{itemize}
    \item The first and second lines contain the kinetic terms of the EFT particles, determined by
    \begin{equation}
        g = \delta_{\overbar{a}b} \, \phi^{\overbar{a} \dagger} \phi^b  + \ldots \, , \quad
        k_{\overbar{p}q} = \delta_{\overbar{p}q} + \ldots \, , \quad
        h_{AB} = \frac{1}{4e^2} \, \delta_{AB} + \ldots \, ,
    \end{equation}
    where the last is holomorphic. We require that $g_{\overbar{a}b}$, $k_{\overbar{p}q}$ and $\re h_{AB}$ be positive definite.\footnote{Some operators contributing to these functions could have been generated from $X S_i$ instead and hence reassigned to the fifth line, but we choose not to do so. We also take $g$ to be real, $k_{\overbar{p} q} = (k_{\overbar{q} p})^{\dagger}$ and $h_{AB} = h_{BA}$. The gauge coupling is denoted as $e$.}
    
    \item The third line contains the two- and four-fermion operators including fermion masses, given by $m_{pr}$ and $r_{\overbar{p} q \overbar{r} s}$ where the former is holomorphic.\footnote{We take $m_{pr} = m_{rp}$ and $r_{\overbar{p} q \overbar{r} s} = r_{\overbar{p} s \overbar{r} q} = r_{\overbar{r} q \overbar{p} s}$.}
    
    \item The fourth and fifth lines contain the auxiliary field terms. The couplings between the auxiliary fields $F$ and $D^A$ are given by a symmetric non-holomorphic function
    \begin{equation}
        \zeta_{AB} = c_{\mathcal{X}a} \left [ \frac{1}{2} \, \delta_{AB} \, \phi^{\overbar{a} \dagger} \phi^a \right ] + \ldots \, .
    \end{equation}
    This arises from the supersymmetrization via $\overbar{X} W^{A\alpha} W_\alpha^B$ of non-holomorphic gauge operators $c_{\text{nh}} O_{\text{nh}}$ like $O_{\mathcal{X}a} = \tr \, [F^{\mu\nu} F_{\mu\nu}^-] \, \phi^{\overbar{a} \dagger} \phi^a$. The linear terms in $F^a$ and $F^p$ are given by holomorphic functions $w$ and $f_p$ subject to the conditions\footnote{They are automatically satisfied if the corresponding $\phi^a$ or $\psi^p$ is charged. We do not introduce any Fayet–Iliopoulos term in $p_A$.}
    \begin{equation}
        w_{,a}(0) = 0 \, , \quad f_p(0) = 0 \, ,
    \end{equation}
    so that only $F$ acquires a vev, while the linear term in $D^A$ is given by a real function
    \begin{equation}
        p_A = - \sum_a g_{,a} \, T_A^{\,a} \, \phi^a \, .
    \end{equation}
    
    \item The sixth line contains the Goldstino terms, which are all $\mathcal{O}(f^{-1})$ other than the dimension four kinetic term; there is no order $f$ Goldstino coupling in $X S_i \, \vbar_{\theta^2 \overbar{\theta}^2} \subset \mathcal{L}_X$ and any Goldstino coupling in $\mathcal{L}_{\text{non-}X}$ is automatically subleading.
\end{itemize}

We now integrate out the auxiliary fields so that only the EFT particles and the Goldstino remain in the theory. The fourth and fifth lines, together with any $\mathcal{O}(f^{-1})$ couplings that include the Goldstino, yield the equations of motion
\begin{align}
    \overbar{F} + \frac{1}{f} \, \zeta_{AB} \, D^A D^B + f - \frac{1}{f} \sum_i c_i O_i &= \mathcal{O} \left ( \frac{1}{f^2} \right ) \enspace \text{and} \enspace \text{h.c.} \, , \\
    g_{\overbar{a}b} \, F^{\overbar{a} \dagger} + w_{,b} &= \mathcal{O} \left ( \frac{1}{f} \right ) \enspace \text{and} \enspace \text{h.c.} \, , \\
    k_{\overbar{p}q} \, F^{\overbar{p} \dagger} + f_q &= \mathcal{O} \left ( \frac{1}{f} \right ) \enspace \text{and} \enspace \text{h.c.} \, , \\
    2 D^B \left [ h_{AB} + \frac{1}{f} \, \zeta_{AB} \, F + \text{c.c.} \right ] - p_A &= \mathcal{O} \left ( \frac{1}{f} \right ) \, .
\end{align}
Note that the first equation arising from $\delta \mathcal{L} / \delta F$ holds up to $\mathcal{O}(f^{-2})$.\footnote{Any term in $\mathcal{L}$ with derivatives on $F$ also contains other fields, and is already $\mathcal{O}(f^{-2})$.} Denoting inverses by upper indices and writing $h'_{AB} = h^{\vphantom{'}}_{AB} - \zeta^{\vphantom{'}}_{AB}$, we can solve these equations as series in $f$ to get
\begin{align}
    \overbar{F} &= - \, f + \frac{1}{f} \left [ \sum_i c_i O_i - \frac{\zeta_{AB}}{16} \, (\re h')^{AC} \, (\re h')^{BD} \, p_C \, p_D \right ] + \mathcal{O} \left ( \frac{1}{f^2} \right ) \enspace \text{and} \enspace \text{h.c.} \, , \\
    F^{\overbar{a} \dagger} &= - \, g^{b\overbar{a}} \, w_{,b} + \mathcal{O} \left ( \frac{1}{f} \right ) \enspace \text{and} \enspace \text{h.c.} \, , \\
    F^{\overbar{p} \dagger} &= - \, k^{q\overbar{p}} \, f_q + \mathcal{O} \left ( \frac{1}{f} \right ) \enspace \text{and} \enspace \text{h.c.} \, , \\
    D^B &= \frac{1}{4} \, (\re h')^{BA} \, p_A + \mathcal{O} \left ( \frac{1}{f} \right ) \, .
\end{align}
Setting the auxiliary fields to their solutions finally yields
\begin{align}
    \mathcal{L}_{\text{aux}+G} &\rightarrow - f^2 + \left [ \sum_i c_i O_i + \text{c.c.} \right ] - v - \frac{i}{2} \, \overbar{G} \overbar{\sigma}^\mu \overset{\text{\tiny$\leftrightarrow$}}{\partial}_\mu G + \mathcal{O} \left ( \frac{1}{f} \right ) \, ,
\end{align}
giving rise to a scalar potential
\begin{equation}
    v = g^{b\overbar{a}} \, w_{,b\vphantom{\overbar{a}}} \, {w_{,\overbar{a}}}^{\dagger} + k^{q\overbar{p}} \, f_q \, {f_{\overbar{p}}}^{\dagger} + \frac{1}{8} \, (\re h')^{BA} \, p_B \, p_A \, , \label{eq:scalar_potential_condition}
\end{equation}
along with any $O_i$ embedded in $S_i$. These account for the EFT operators missing from \cref{eq:direct_SUSY_operators}. Moreover, no Goldstino interaction at order $f^0$ or above is generated in the process, ensuring that this additional degree of freedom introduced to supersymmetrize the EFT remains hidden at leading order.

Working explicitly to dimension six, we have been able to construct a class of nonlinearly supersymmetric theories $\mathcal{L} = \mathcal{L}_X + \mathcal{L}_{\text{non-}X}$ that approximates most EFTs arbitrarily closely. Indeed, all operators containing at least one fermion or gauge boson can be supersymmetrized cleanly in $\mathcal{L}_X$ via the SUSY breaking auxiliary field $F$, such that any additional terms required to realize SUSY are Goldstino couplings suppressed by $f$. Nevertheless, the scalar sector is not obviously amenable to this scheme; our choice to derive it from a nonlinear sigma model $\mathcal{L}_{\text{non-}X}$ of constrained superfields means that restrictions akin to unconstrained ones remain:
\begin{itemize}
    \item As in \cref{eq:scalar_kinetic_condition}, the scalar kinetic term $g_{\overbar{a}b}$ must be Kähler.
    \item As in \cref{eq:scalar_potential_condition}, the scalar potential $v$ must be a sum of the usual $F^a$-, $F^p$- and $D^A$-terms, but with a modified $h'$ resulting from the supersymmetrization of non-holomorphic gauge operators. In particular, the tunable inputs $w$ and $f_p$ must be holomorphic (and gauge covariant).\footnote{Other inputs to \cref{eq:scalar_potential_condition} are fixed by other EFT operators that must be supersymmetrized.}
\end{itemize}
These requirements on $\phi$-dependent functions span scalar operators across different mass dimensions, and apply beyond dimension six even if other higher dimensional operators can be supersymmetrized via $F$. However, this does not preclude the possibility for more general strategies that circumvent these limitations, such as allowing auxiliary field terms of higher degrees, imposing alternative superfield constraints, or introducing further degrees of freedom beyond the Goldstino and the current set of auxiliary fields. We leave their exploration to future work, and turn to the geometric applications that the present supersymmetrization scheme brings to the broad range of EFTs it already encompasses.

\section{Geometry of the superfield target space}
\label{sec:geometry}

Having shown how it is possible to supersymmetrize a wide class of EFTs, we now give a reason why it is useful to do so, whether or not SUSY should be present on other grounds. While minimal modifications have been made to the theory when introducing the Goldstino and auxiliary fields, they elucidate new structures among EFT operators under field redefinitions that would not be apparent otherwise. By packaging the original particles and additional components into constrained superfields, we can enable and untangle redefinitions between fields of various spins using the properties of superspace, systematically organizing operators in the process into geometric objects on the target space of superfields. With attention to the gauge sector, we proceed to establish a vector bundle geometry for the field strength superfield $W_\alpha$ and an underlying geometry stemming from the vector superfield $V$. Along the way, we extend the literature by incorporating both CP-odd and even operators via complex geometry, and enacting a range of redefinitions across different spins. We finally formulate a geometrically covariant derivative for the $X S_i$ couplings previously used in supersymmetrizing the EFT.

\subsection{A holomorphic vector bundle}

To put the superfield theory in geometric form, we set aside the $X S_i$ couplings for now and take apart each matter representation into its components. Writing $\Phi^{I_a}$ for the $(\mathrm{dim} \, R^a)$ components of $Z^a$ and similarly $\Phi^{I_p}$ for $Y^p$, we assemble them along with $X$ into
\begin{equation}
    \{ \Phi^I \} = \{ X \} \cup \{ \Phi^{I_a} \text{ for all } a \} \cup \{ \Phi^{I_p} \text{ for all } p\} \, .
\end{equation}
These become coordinates for a complex manifold $\mathcal{M}$, allowing for holomorphic redefinitions
\begin{equation}
    \Phi^I = \Phi^I(\Phi^{J'}) \, .
\end{equation}
Operators in the scalar and fermion sectors collect into a Kähler potential $K(\overbar{\Phi}, \Phi)$ and a superpotential $W(\Phi)$ on $\mathcal{M}$, along with the minimal Goldstino terms.\footnote{Namely, the Komargodski-Seiberg action \cite{Komargodski:2009rz}. Strictly speaking $\mathcal{M}$ may not be Kähler, but we will make this reasonable assumption in view of \cref{eq:metric}.} Meanwhile, the components $W_\alpha^A$ are coordinates for a manifold of dimension $(\mathrm{dim} \, \mathfrak{g})$, and chart a total space $\mathcal{W}$ when combined with $\Phi^I$. To incorporate operators in the gauge sector geometrically, additional structure is imposed on the total space. As a starting point, we demand that $\mathcal{W}$ forms a holomorphic vector bundle over $\mathcal{M}$, thereby restricting to redefinitions\footnote{We will see shortly how these can arise from a complex reparameterization of the gauge bosons. In particular, $\supset$ indicates that there can be other $V$-dependent terms in the redefinitions invisible to $\mathcal{W}$.}
\begin{equation}
    W_\alpha^A \supset \tensor{\tau}{^A_{A'}}(\Phi) \, W_\alpha^{A'} \, , \label{eq:W_redef}
\end{equation}
where $\tensor{\tau}{^A_{A'}}$ is an invertible complex matrix, and elevating the gauge kinetic function $h_{AB}(\Phi)$ to a symmetric bilinear form on $\mathcal{W}$.

These geometric quantities on $\mathcal{M}$ and $\mathcal{W}$ determine the non-derivative superfield couplings
\begin{equation}
    \mathcal{L}_{\text{non-deriv}} = \int d^4 \theta \; K(\overbar{\Phi}, e^{2V} \Phi) + \left \{ \int d^2 \theta \; \Big [ \, W(\Phi) + h_{AB}(\Phi) \, W^{\alpha A} W_\alpha^B \, \Big ] + \text{c.c.} \right \} \subset \mathcal{L} \, ,
\end{equation}
subject to gauge invariance. On $\mathcal{M}$, gauge symmetry is encoded in holomorphic Killing vectors
\begin{equation}
    v^I_A(\Phi) = \begingroup
    \renewcommand\arraystretch{1.2}
    \begin{pmatrix}
        0 \\
        i \, T^a_A Z^a \\
        i \, T^p_A Y^p \\
    \end{pmatrix}
    \endgroup \, ,
\end{equation}
which preserve the two potentials under the action
\begin{equation}
    \delta_\Omega \Phi^I = - 2 \, \Omega^A \, v^I_A(\Phi) \, ,
\end{equation}
and yield the real momentum maps $p_A = i v^I_A K_{,I}^{\vphantom{I}}$.\footnote{The Killing vectors are presented in the block form $\{X, Z^a, Y^p\}$ and happen to be linear in the unprimed coordinates of the previous section. More generally, $K$ is allowed to change by a Kähler transformation under the gauge symmetry.} On $\mathcal{W}$, we require
\begin{equation}
    v^I_C \, h_{AB,I}^{\vphantom{I}} = \tensor{f}{_{CA}^D} \, h_{DB}^{\vphantom{C}} + \tensor{f}{_{CB}^D} \, h_{AD}^{\vphantom{C}} \, , \label{eq:h_invariance}
\end{equation}
i.e. the bilinear form resides in the direct product of adjoint representations.

There are similarities and differences between the geometric roles that the form $h_{AB}$ plays on $\mathcal{W}$, and the metric
\begin{align}
    g_{\overbar{I}J} &\equiv K_{,\overbar{I}J} \label{eq:metric} \\[2pt]
    &= \begingroup
    \setlength\arraycolsep{5pt}
    \renewcommand\arraystretch{1.2}
    \begin{pmatrix}
        1 & 0 & 0 \\
        0 & g_{\overbar{a}b} + Y^{\overbar{p} \dagger} \, k_{\overbar{p}q, \overbar{a}b} \, Y^q + Y^{\overbar{p} \dagger} \, Y^{\overbar{r} \dagger} \, r_{\overbar{p} q \overbar{r} s, \overbar{a} b} \, Y^q \, Y^s & Y^{\overbar{p} \dagger} \, k_{\overbar{p}q, \overbar{a}} + 2 \, Y^{\overbar{p} \dagger} \, Y^{\overbar{r} \dagger} \, r_{\overbar{p} q \overbar{r} s, \overbar{a}} \, Y^s \\
        0 & k_{\overbar{p}q,b} \, Y^q + 2 \, Y^{\overbar{r} \dagger} \, r_{\overbar{p} q \overbar{r} s, b} \, Y^q \, Y^s & k_{\overbar{p}q} + 4 \, Y^{\overbar{r} \dagger} \, r_{\overbar{p} q \overbar{r} s} \, Y^s \\
    \end{pmatrix}
    \endgroup \, , \notag
\end{align}
on $\mathcal{M}$. The latter contains the positive definite kinetic terms of the Goldstino, scalars and fermions, and can hence be inverted to give a connection compatible with the complex structure and the metric, with coefficients
\begin{equation}
    \tensor{\Gamma}{^I_{KJ}} = g^{I\overbar{L}} \, g_{\overbar{L}J,K} \, . \label{eq:g_connection}
\end{equation}
The connection enables covariant differentiation and endows $\mathcal{M}$ with curvature
\begin{equation}
    R_{\overbar{I} J \overbar{K} L} = g_{\overbar{I}N}^{\vphantom{N}} \, \tensor{\Gamma}{^N_{LJ, \overbar{K}}} = g_{\overbar{I}J, \overbar{K}L} - g_{\overbar{I}N, \overbar{K}} \, g^{N\overbar{M}} \, g_{\overbar{M}J, L} \, .
\end{equation}
Similarly, the real part of $h_{AB}$ is the positive definite kinetic term of the gauge bosons, so that the full form is invertible and has a symmetric square root $(\sqrt{h})_{AB}$.\footnote{Defined where the square root is analytic in $\Phi^I$. A complex matrix whose real part is positive definite is sometimes also termed positive definite, even if it is not Hermitian.} A connection given by e.g.
\begin{equation}
    \tensor{\Gamma}{^A_{IB}} = (\sqrt{h})^{AC} \, (\sqrt{h})_{CB,I} \, , \label{eq:h_connection}
\end{equation}
would be compatible with both the holomorphic structure and the symmetric form on $\mathcal{W}$. However, this connection is not the unique possibility unlike a Hermitian form, and its curvature actually vanishes. Still, it will be useful for formulating a covariant derivative later.\footnote{A non-compatible connection can alternatively be used like in \cite{Assi:2023zid, Derda:2024jvo, Assi:2025fsm, Craig:2025uoc}, although there is no distinguished choice here.}

When $\tensor{\tau}{^A_{A'}}$ is purely $Z$-dependent, \cref{eq:W_redef} enacts the component redefinitions
\begin{align}
    F_{\mu\nu}^A &\supset \tensor{\tau}{^A_{A'}}(\phi) \, F_{\mu\nu}^{A'} + \mathcal{O} \left ( \frac{1}{f} \right ) \, , \\
    D^A &\supset \tensor{\tau}{^A_{A'}}(\phi) \, D^{A'} + \mathcal{O} \left ( \frac{1}{f} \right ) \, ,
\end{align}
at leading order in $f$ within the EFT. On a real vector bundle, the real and imaginary parts of $h_{AB}(\phi)$ entering
\begin{equation}
    \re h_{AB} \, \Big [ - F^{\mu\nu A} F_{\mu\nu}^B + 2 D^A D^B \, \Big ] + \im h_{AB} \, \Big [ - F^{\mu\nu A} \widetilde{F}_{\mu\nu}^B \, \Big ] \subset \mathcal{L}_{\text{non-deriv}} \, ,
\end{equation}
would be regarded as two separate entities, with the former assuming a special role as an invertible bundle metric. By allowing $\tau$ to be complex, we incorporate the latter as well into the distinguished geometry of $\mathcal{W}$. The inclusion of both CP-odd and even gauge operators in a complex form becomes natural upon supersymmetrization.

\subsection{From the vector superfield to field strength}

The redefinitions of $W_\alpha$ above can be implemented by redefinitions
\begin{equation}
    V^A = \tensor{\rho}{^A_{A'}}(\overbar{\Phi}, \Phi) \, V^{A'} \, , \label{eq:V_redef}
\end{equation}
of the underlying $V$. Taking $\tensor{\rho}{^A_{A'}}$ to be an invertible complex matrix, these are the transition functions of a complex vector bundle $\mathcal{V}$ over $\mathcal{M}$ of the same rank as $\mathcal{W}$. Forming the jet bundle $\mathcal{J}^r$ with the superspace manifold $\mathcal{T}$ as the source and $\mathcal{V}$ as the target, we then obtain $W_\alpha^A$ as an object on $\mathcal{J}^3$ where $r=3$ is the jet order.\footnote{The order of a jet counts the number of source derivatives on the target coordinates.} From the chiral identities
\begin{equation}
    \overbar{D}^2 D_\alpha \overbar{\Phi}^{\overbar{I}} = 0 \enspace \text{and} \enspace \overbar{D}^2 D_\alpha \Phi^I = 0 \, ,
\end{equation}
we find
\begin{equation}
    W_\alpha^A = - \frac{1}{8} \, \overbar{D}^2 \left ( e^{-2V} D_\alpha \, e^{2V} \right )^A = \left [ \tensor{\rho}{^A_{A'}} + \mathcal{O}(V) \right ] W_\alpha^{A'} + (r \leq 2) \, ,
\end{equation}
so that the vector bundle structure of $\mathcal{V}$ persists on the $V = 0$ slice of its jet bundle, reproducing $\mathcal{W}$ upon comparison with \cref{eq:W_redef} for holomorphic $\rho$. In words, $\mathcal{W}$ is the geometry obtained from $\mathcal{V}$ over its null section when focusing on the top jet order.

There is however more structure to be found in the geometry of $\mathcal{V}$ than $\mathcal{W}$. Similar to the construction of Lagrange spaces \cite{Craig:2023wni, Miron:1997vjh}, we can rewrite the chiral super-operators in terms of $V$ to get
\begin{align}
    \mathcal{L}_{\text{chiral}} &= \int d^2 \theta \; \Big [ \, W(\Phi) + h_{AB}(\Phi) \, W^{\alpha A} W_\alpha^B \, \Big ] \\
    &= \int d^2 \theta \; \Big [ \, \frac{1}{16} \, \eta_{AB}(\Phi, V) \, \left ( \overbar{D}^2 D^\alpha V^A \right ) \left ( \overbar{D}^2 D_\alpha V^B \right ) + (r \leq 2) \, \Big ] \, ,
\end{align}
where
\begin{equation}
    \eta_{AB}(\Phi, V) = \Big [ \delta_A^C + i \, \tensor{f}{_{AE}^C} \, V^E - \frac{2}{3} \, \tensor{f}{_{AE}^G} \tensor{f}{_{GF}^C} \, V^E V^F \Big ] \, h_{CD}(\Phi) \, \Big [ (A,C) \leftrightarrow (B,D) \Big ] \, ,
\end{equation}
is now $V$-dependent. Under \cref{eq:V_redef}, the transformation
\begin{equation}
     \eta_{A'B'}^{\vphantom{A}} = \tensor{\rho}{^A_{A'}} \, \eta_{AB}^{\vphantom{A}} \, \tensor{\rho}{^B_{B'}} \, , \label{eq:eta_transf}
\end{equation}
entails that $\eta_{AB}$ is a symmetric bilinear form that is analogous to $h_{AB}$ but varies along the fibers of the bundle. One can regard $\eta_{AB}$ as an extension of $h_{AB}$ to the total space of $\mathcal{V}$ that carries further information on the algebraic structure of the gauge group.\footnote{To get a bilinear form on the total space, we must modify the implicit $dV^A \otimes dV^B$ to
\begin{equation}
    \delta V^A = dV^A + \tensor{\Gamma}{^A_{IB \vphantom{\overbar{I}}}} \, V^B \, d\Phi^I + \tensor{\Gamma}{^A_{\overbar{I}B}} \, V^B \, d\overbar{\Phi}^{\overbar{I}} \, ,
\end{equation} using a connection like \cref{eq:h_connection} built from $h$, i.e. the value of $\eta$ on the null section $V = 0$ only. In unprimed coordinates, the connection coefficients $\tensor{\Gamma}{^A_{\overbar{I}B}} = 0$ happen to vanish along the anti-holomorphic directions on $\mathcal{M}$. On the other hand, $\partial / \partial V^A$ requires no modification.} While bearing no new information, the Kähler potential on $\mathcal{M}$ can also be extended to the total space as\footnote{Note that $V^A V^B V^C = 0$ due to the constraint \cref{eq:V_constraint}.}
\begin{equation}
    K(\overbar{\Phi}, e^{2V} \Phi) = K(\overbar{\Phi}, \Phi) - 2 \, p_A(\overbar{\Phi}, \Phi) \, V^A + 2 \, g_{\overbar{I}J}(\overbar{\Phi}, \Phi) \, \overbar{v}^{\overbar{I}}_A(\overbar{\Phi}) \, v^J_B(\Phi) \, V^A V^B \, , \label{eq:K_on_W}
\end{equation}
where each of $K$, $p_A$ and $g_{\overbar{I}J} \overbar{v}^{\overbar{I}}_A v^J_B$ is moreover covariant on $\mathcal{V}$. The gauge symmetry of these extended quantities can be specified by likewise augmenting the Killing vectors according to
\begin{equation}
    \delta_\Omega V = i \, \mathrm{ad}_V \, (\Omega + \Omega^\dagger) + i \, \mathrm{ad}_V \, \mathrm{coth} \; \mathrm{ad}_V \, (\Omega - \Omega^\dagger) \, ,
\end{equation}
where $\mathrm{ad}_V \, \Omega = [V, \Omega]$ and $\mathrm{coth}$ is implemented as a power series.\footnote{To be clear, gauge transformations are vector flows on $\mathcal{V}$ and not coordinate changes.}

To translate the redefinitions into component form, $V^{B'}$ can be expanded like in \cref{eq:V} but with generically complex components. When $\tensor{\rho}{^A_{A'}}$ is purely $Z$-dependent, one finds at leading order within the EFT that \cref{eq:V_redef} enacts\footnote{An example of complex $A_\mu^{B'}$ comes from the weak bosons $W_\mu^\pm$.}
\begin{align}
    A_\mu^B &= \tensor{\rho}{^B_{B'}}(\overbar{\phi}, \phi) \, A_\mu^{B'} + \mathcal{O} \left ( \frac{1}{f} \right ) \, , \label{eq:A_redef} \\
    D^B &= \tensor{\rho}{^B_{B'}} \, D^{B'} - i \, \tensor{\rho}{^B_{B',\overbar{b}}} \; A_\mu^{B'} \, \partial^\mu \phi^{\overbar{b} \dagger} + i \, \tensor{\rho}{^B_{B',b \vphantom{\overbar{b}}}} \; A_\mu^{B'} \, \partial^\mu \phi^b + \mathcal{O} \left ( \frac{1}{f} \right ) \, ,
\end{align}
while the other components remain suppressed by $f$. Redefinitions of the field strength follow from the first, and the auxiliary field remains non-propagating under the second. Both generate additional derivative couplings of scalar fields at the same jet order as those of the gauge bosons, such as
\begin{equation}
    (\re h_{BC}) \, \tensor{\rho}{^B_{B',\overbar{b}}} \, \tensor{\rho}{^C_{C',c \vphantom{\overbar{b}}}} \left [ - \, 2 \, \eta_{\mu\nu}^{\vphantom{B'}} \, A_\sigma^{B'} A^{\sigma C'} + 4 \, A_\mu^{B'} A_\nu^{C'} + (\mu \leftrightarrow \nu) \, \right ] \partial^\mu \phi^{\overbar{b} \dagger} \, \partial^\nu \phi^c \, .
\end{equation}
That the top order isolates the field strength redefinition is a neat feature of the superfield description, disentangling the gauge bosons from the scalars and enabling straightforward geometric formulations for both.

\subsection{From vector to fiber bundles}

Having established a vector bundle geometry arising from the gauge kinetic function, we can now relax the structure imposed on the total space to accommodate a larger set of field redefinitions. Holding redefinitions of $\Phi^I$ to be $V$-independent, we can extend \cref{eq:V_redef} to
\begin{equation}
    V^A = \varrho^A(\overbar{\Phi}, \Phi) + \tensor{\rho}{^A_{A'}}(\overbar{\Phi}, \Phi) \, V^{A'} \, ,
\end{equation}
where $\varrho^A$ is a $V$-independent shift in the coordinate origin, making $\mathcal{V}$ an affine bundle modeled after the vector bundle. For suitable $\varrho^A$, this implements
\begin{equation}
    A_\mu^B = - \, i \, \tensor{\varrho}{^B_{,\overbar{b}}} \; \partial_\mu \phi^{\overbar{b} \dagger} + i \, \tensor{\varrho}{^B_{,b \vphantom{\overbar{b}}}} \; \partial_\mu \phi^b - \tensor{\varrho}{^B_{,\overbar{p}q}} \; (\psi^{\overbar{p} \dagger} \overbar{\sigma}_\mu \psi^q) + \cref{eq:A_redef} \, ,
\end{equation}
which now involves redefinitions across strictly different spin.\footnote{However, the other components of $V^{A'}$ may no longer be suppressed by $f$.} More generally, we allow the redefinitions
\begin{align}
    V^A &= V^A(\overbar{\Phi}, \Phi, V') \\
    &= \varrho^A(\overbar{\Phi}, \Phi) + \tensor{\rho}{^A_{A'}}(\overbar{\Phi}, \Phi) \, V^{A'} + \tensor{P}{^A_{B'C'}}(\overbar{\Phi}, \Phi) \, V^{B'} V^{C'} + \ldots \, ,
\end{align}
enabling $D^A \supset \tensor{P}{^A_{B'C'}}(\overbar{\phi}, \phi) \, A_\mu^{B'} A^{\mu C'}$ in particular. On what is now a generic fiber bundle $\mathcal{V}$, some geometric properties from before continue to apply, namely that \cref{eq:eta_transf} holds with $\tensor{\rho}{^A_{A'}}$ replaced by $\partial V^A / \partial V^{A'}$ and the inverse form $\eta^{AB}$ remains covariant.

\subsection{Derivative couplings}

We now restore the $X S_i$ couplings employed in the previous section to supersymmetrize most of the EFT operators. Other than $\overbar{X} Y^p Y^q$ which fits into the Kähler potential, the rest are derivative couplings containing $\nabla_\alpha$ from \cref{eq:gauge_chiral_deriv}, which is gauge and chiral covariant but not always field redefinition covariant. For redefinitions on $\mathcal{M}$ and the vector bundle $\mathcal{W}$, the geometric connections above can be used to promote $\nabla_\alpha$ into a derivative $\widetilde{\nabla}_\alpha$ that attains all three types of covariance.

For $\Phi^I = X$, $Z^a$ and $Y^p$, the first derivative remains unchanged because
\begin{equation}
    \widetilde{\nabla}_\alpha \Phi^I \equiv \nabla_\alpha \Phi^I = D_\alpha \Phi^I - i \left ( e^{-2V} D_\alpha \, e^{2V} \right )^A v^I_A(\Phi) \, ,
\end{equation}
is already field redefinition covariant on $\mathcal{M}$. While higher order derivatives have not been used at dimension six, they should in any case be modified using the connection \cref{eq:g_connection}. At second order, we have
\begin{equation}
    \widetilde{\nabla}_\beta \widetilde{\nabla}_\alpha \Phi^I \equiv \left [ \delta^I_J \, D_\beta - i \left ( e^{-2V} D_\beta \, e^{2V} \right )^A v^I_{A,J} + \tensor{\Gamma}{^I_{KJ}} \left ( \widetilde{\nabla}_\beta \Phi^K \right ) \right ] \widetilde{\nabla}_\alpha \Phi^J \, ,
\end{equation}
and subsequent orders follow from compounding the formula. For $W_\alpha$, we use the connection \cref{eq:h_connection} to set\footnote{The appearances of $V$ do not spoil covariance on $\mathcal{W}$ because they arise at lower jet order.}
\begin{align}
    \widetilde{\nabla}^{\vphantom{A}}_\beta W_\alpha^A &\equiv \left [ \delta^A_B \, \nabla^{\vphantom{A}}_\beta + \tensor{\Gamma}{^A_{IB}} \left ( \widetilde{\nabla}^{\vphantom{A}}_\beta \Phi^I \right ) \right ] W_\alpha^B \\
    &= D^{\vphantom{A}}_\beta W_\alpha^A + \left \{ \left ( e^{-2V} D_\beta \, e^{2V} \right ) , W_\alpha \right \}^A + \tensor{\Gamma}{^A_{IB}} \left ( \widetilde{\nabla}^{\vphantom{A}}_\beta \Phi^I \right ) W_\alpha^B \, ,
\end{align}
which can again be compounded for higher order derivatives. To get a minimal geometric promotion of $X S_i$, we extract all factors of $W_\alpha^A$, $\widetilde{\nabla}_\alpha \Phi^I$ and $\widetilde{\nabla}^{\vphantom{A}}_\beta W_\alpha^A$ into a new $\widetilde{S}_i$, contract the free $I$ and $A$ indices with a $\Phi$-dependent $\chi_i$, and insert suitable factors of $e^{2V}$. Then $\chi_i$ is a tensor on $\mathcal{M}$ or $\mathcal{W}$ that combines the original EFT operator embedded in $X S_i$ with higher dimensional ones related by field redefinitions.

Meanwhile, there is no obvious derivative that respects both gauge and field redefinition covariance on the vector bundle $\mathcal{V}$ since $V$ is not gauge invariant. The full $\chi_i \widetilde{S}_i$ can always be regarded as a scalar function on the jet bundle $\mathcal{J}^r$. But like the chiral super-operators, we can also expand the non-chiral ones
\begin{equation}
    \mathcal{L}_{\text{non-chiral}} = \int d^4 \theta \; \bigg \{ \cref{eq:K_on_W} + \Big [ \, \sum_i \chi_i \widetilde{S}_i \, (\overbar{\Phi}, \Phi, V, \overbar{D}_{\dot{\alpha}}, D_\alpha) + \text{c.c.} \, \Big ] \bigg \} \, ,
\end{equation}
in derivatives of $\Phi$ and $V$. The $\Phi$- and $V$-dependent functions contracting their indices are objects on the total space of $\mathcal{V}$ that may enjoy special geometric status like $\eta_{AB}$, depending on the derivative structure.

For illustration, a super-operator built from $X Y^{\overbar{p} \dagger} \nabla^\alpha Y^q \nabla_\alpha Z^a$ like \cref{eq:super_operator_example} can be promoted to
\begin{equation}
    \chi_{IJ}(\overbar{\Phi} e^{2V}, \Phi) \, \widetilde{\nabla}^\alpha \Phi^I \widetilde{\nabla}_\alpha \Phi^J \supset \left [ - \, 4 \, \chi_{IJ}(\overbar{\Phi}, \Phi) \, v^I_A(\Phi) \, v^J_B(\Phi) + \mathcal{O}(V) \right ] D^\alpha V^A D_\alpha V^B \, ,
\end{equation}
so that $\chi_{IJ}(\overbar{\Phi}, \Phi)$ is a tensor on $\mathcal{M}$ subject to the gauge symmetry condition
\begin{equation}
    \overbar{v}^{\overbar{K}}_C \, \chi_{IJ,\overbar{K}} + v^K_C \, \chi_{IJ,K \vphantom{\overbar{K}}} = - v^K_{C,I} \, \chi_{KJ \vphantom{\overbar{K}}} - v^K_{C,J} \, \chi_{IK \vphantom{\overbar{K}}} \, .
\end{equation}
In the absence of other derivative couplings, the term prepending $D^\alpha V^A D_\alpha V^B$ collected in square brackets is covariant on the total space of $\mathcal{V}$. As another example, a super-operator built from $\overbar{X} Y^p Y^q (\nabla^\alpha W_\alpha^A)$ can be promoted to
\begin{equation}
    \chi_A(\overbar{\Phi} e^{2V}, \Phi) \, \widetilde{\nabla}^\alpha W_\alpha^A = \left [ \chi_A(\overbar{\Phi}, \Phi) + \mathcal{O}(V) \right ] D^\alpha \overbar{D}^2 D_\alpha V^A + (r \leq 3) \, ,
\end{equation}
so that $\chi_A(\overbar{\Phi}, \Phi)$ is a tensor on $\mathcal{W}$ subject to the gauge symmetry condition
\begin{equation}
    \overbar{v}^{\overbar{K}}_C \, \chi_{A,\overbar{K}} + v^K_C \, \chi_{A,K \vphantom{\overbar{K}}} = \tensor{f}{_{CA}^B} \, \chi_{B \vphantom{\overbar{K}}} \, .
\end{equation}
In the absence of other derivative couplings, the term prepending $D^\alpha \overbar{D}^2 D_\alpha V^A$ collected in square brackets remains as the top jet order under redefinitions on $\mathcal{V}$, and thus can be regarded as a covariant extension of $\chi_A$ to the total space of $\mathcal{V}$.

\section{Conclusion and outlook}
\label{sec:conc}

A supersymmetric lift systematically organizes operators in EFT. We have demonstrated how to supersymmetrize a broad class of EFTs with minimal modifications, by completing particles into constrained superfields and introducing suitably suppressed Goldstino couplings. With operators embedded supersymmetrically, we have then established how they are organized in the vector bundle geometry of superfields under field redefinitions in the gauge sector. SUSY has played an important role in facilitating complex reparameterizations and disentangling spins, elucidating geometric structures that would not be directly obvious otherwise.

Several direct extensions of the present work are possible. Having worked explicitly up to dimension six, the next step on the SUSY front is to apply the supersymmetrization scheme to higher dimensional operators. Alternative schemes that alleviate the current restrictions on supersymmetrizable scalar operators can likewise enlarge the applicability of SUSY to EFT. On the geometry front, more powerful structures arising from e.g. special Kähler manifolds may be uncovered if more SUSY is introduced \cite{Kuzenko:2011ya, Cribiori:2016hdz, Dudas:2017sbi}, and a full treatment of effective gauge theories via the machinery of principal and jet bundles awaits development. These extensions promise to lend deeper insight into the structure and organization underlying EFT operators.

More generally, the supersymmetrization of EFTs can be independently useful beyond field redefinitions, with already a range of theoretical and phenomenological applications \cite{Antoniadis:2010hs, Khoury:2010gb, Dudas:2012fa, Alonso:2014rga, Elias-Miro:2014eia, Cheung:2015aba, Cohen:2016jzp, Cohen:2016dcl, Cohen:2018qvn, Cohen:2019gsc}. While subleading Goldstino operators have not been the present focus, their further study may reveal formal information on the leading operators they originate from, or physical implications should exact SUSY hold. Relatedly, like how approximate SUSY may not be obvious in the EFT at first glance, hidden symmetries in other contexts can lead to intricate and interesting consequences \cite{Craig:2021ksw}.

\acknowledgments

The authors thank I-Kwan Lee and Zhengkang Zhang for helpful conversations, and Benoît Assi, Timothy Cohen and Dave Sutherland for valuable comments on the manuscript. AF and YTL are grateful for support from the Worster Summer Research Fellowship. This work is supported in part by the U.S. Department of Energy under the grant DE-SC0011702 and performed in part at the Kavli Institute for Theoretical Physics, supported by the National Science Foundation under the grant NSF PHY-1748958.

\appendix
\section{Lie algebra and spinor conventions}
\label{app:conv}

Given a Lie group, we follow the convention that the generators $T_A = (T_A)^\dagger$ of its Lie algebra $\mathfrak{g}$ are Hermitian and satisfy
\begin{align}
    \tr \, T_A T_B &= \frac{1}{2} \, \delta_{AB} \, , \\
    \left [ T_A, \, T_B \right ] &= i \, \tensor{f}{_{AB}^C} \, \tensor{T}{_C} \, .
\end{align}
We use the metric signature $\eta_{\mu\nu} = \mathrm{diag}(+, -, -, -)$ and follow the spinor conventions
\begin{alignat}{2}
    \epsilon^{12} &= - \epsilon_{12} = 1 \, , &\epsilon^{0123} &= - \epsilon_{0123} = 1 \, , \\[0.3em]
    \sigma^\mu_{\alpha \dot{\alpha}} &= \left ( \delta^{\vphantom{i}}_{\alpha \dot{\alpha}}, \, \sigma^i_{\alpha \dot{\alpha}} \right ) \, , &\overbar{\sigma}^{\mu \dot{\alpha} \alpha} &= \left ( \delta^{\dot{\alpha} \alpha}, \, - \, \sigma^{i \dot{\alpha} \alpha} \right) \, , \\
    \sigma^{\mu\nu} &= \frac{i}{4} \left ( \sigma^\mu \overbar{\sigma}^\nu - \sigma^\nu \overbar{\sigma}^\mu \right ) \, , \qquad &\overbar{\sigma}^{\mu\nu} &= \frac{i}{4} \left ( \overbar{\sigma}^\mu \sigma^\nu - \overbar{\sigma}^\nu \sigma^\mu \right ) \, ,
\end{alignat}
where $\sigma^i / 2$ are the three generators of $\mathfrak{su}(2)$.

\bibliographystyle{JHEP}
\bibliography{arXiv_v1}

@inproceedings{Oppenheimer:1956hfa,
    author = "Oppenheimer, J. R. and Yang, C. N. and Wentzel, C. and Marshak, R. E. and Dalitz, R. H. and Gell-Mann, M. and Markov, M. A. and D'Espagnat, B.",
    title = "{Theoretical interpretation of new particles}",
    booktitle = "{6th Annual Rochester Conference on High energy nuclear physics}",
    pages = "VIII.1--36",
    year = "1956"
}

@article{Lee:2024xqa,
    author = "Lee, Yu-Tse",
    title = "{Field space geometry and nonlinear supersymmetry}",
    eprint = "2410.21395",
    archivePrefix = "arXiv",
    primaryClass = "hep-th",
    doi = "10.1103/PhysRevD.111.105004",
    journal = "Phys. Rev. D",
    volume = "111",
    number = "10",
    pages = "105004",
    year = "2025"
}

@article{Delgado:2022bho,
    author = "Delgado, Antonio and Martin, Adam and Wang, Runqing",
    title = "{Constructing operator basis in supersymmetry: a Hilbert series approach}",
    eprint = "2212.02551",
    archivePrefix = "arXiv",
    primaryClass = "hep-th",
    doi = "10.1007/JHEP04(2023)097",
    journal = "JHEP",
    volume = "04",
    pages = "097",
    year = "2023"
}

@article{Delgado:2023ivp,
    author = "Delgado, Antonio and Martin, Adam and Wang, Runqing",
    title = "{Counting operators in N = 1 supersymmetric gauge theories}",
    eprint = "2305.01736",
    archivePrefix = "arXiv",
    primaryClass = "hep-th",
    doi = "10.1007/JHEP07(2023)081",
    journal = "JHEP",
    volume = "07",
    pages = "081",
    year = "2023"
}

@article{Delgado:2023ogc,
    author = "Delgado, Antonio and Martin, Adam and Wang, Runqing",
    title = "{Hidden U(N) symmetry behind $ \mathcal{N} $ = 1 superamplitudes}",
    eprint = "2309.15802",
    archivePrefix = "arXiv",
    primaryClass = "hep-th",
    doi = "10.1007/JHEP11(2023)215",
    journal = "JHEP",
    volume = "11",
    pages = "215",
    year = "2023"
}

@article{Delgado:2024ivu,
    author = "Delgado, Antonio and Martin, Adam and Wang, Runqing",
    title = "{Basis for non-factorizable superamplitudes in $ \mathcal{N} $ = 1 supersymmetry}",
    eprint = "2406.01861",
    archivePrefix = "arXiv",
    primaryClass = "hep-th",
    doi = "10.1007/JHEP09(2024)051",
    journal = "JHEP",
    volume = "09",
    pages = "051",
    year = "2024"
}

@article{Delgado:2025oev,
    author = "Delgado, Antonio and Martin, Adam and Wang, Runqing",
    title = "{Non-factorizable superamplitudes for massive $ \mathcal{N} $ = 1 superstates}",
    eprint = "2505.08741",
    archivePrefix = "arXiv",
    primaryClass = "hep-th",
    doi = "10.1007/JHEP08(2025)201",
    journal = "JHEP",
    volume = "08",
    pages = "201",
    year = "2025"
}

@article{Elias-Miro:2014eia,
    author = "Elias-Miro, J. and Espinosa, J. R. and Pomarol, A.",
    title = "{One-loop non-renormalization results in EFTs}",
    eprint = "1412.7151",
    archivePrefix = "arXiv",
    primaryClass = "hep-ph",
    doi = "10.1016/j.physletb.2015.05.056",
    journal = "Phys. Lett. B",
    volume = "747",
    pages = "272--280",
    year = "2015"
}

@article{Volkov:1973ix,
    author = "Volkov, D. V. and Akulov, V. P.",
    title = "{Is the Neutrino a Goldstone Particle?}",
    doi = "10.1016/0370-2693(73)90490-5",
    journal = "Phys. Lett. B",
    volume = "46",
    pages = "109--110",
    year = "1973"
}

@article{Ivanov:1978mx,
    author = "Ivanov, E. A. and Kapustnikov, A. A.",
    title = "{General Relationship Between Linear and Nonlinear Realizations of Supersymmetry}",
    reportNumber = "JINR-P2-11514",
    doi = "10.1088/0305-4470/11/12/005",
    journal = "J. Phys. A",
    volume = "11",
    pages = "2375--2384",
    year = "1978"
}

@article{Rocek:1978nb,
    author = "Rocek, M.",
    title = "{Linearizing the Volkov-Akulov Model}",
    doi = "10.1103/PhysRevLett.41.451",
    journal = "Phys. Rev. Lett.",
    volume = "41",
    pages = "451--453",
    year = "1978"
}

@article{Lindstrom:1979kq,
    author = "Lindstrom, U. and Rocek, M.",
    title = "{CONSTRAINED LOCAL SUPERFIELDS}",
    doi = "10.1103/PhysRevD.19.2300",
    journal = "Phys. Rev. D",
    volume = "19",
    pages = "2300--2303",
    year = "1979"
}

@article{Casalbuoni:1988xh,
    author = "Casalbuoni, R. and De Curtis, S. and Dominici, D. and Feruglio, F. and Gatto, Raoul",
    title = "{Nonlinear Realization of Supersymmetry Algebra From Supersymmetric Constraint}",
    reportNumber = "UGVA-DPT-1988/12-601",
    doi = "10.1016/0370-2693(89)90788-0",
    journal = "Phys. Lett. B",
    volume = "220",
    pages = "569--575",
    year = "1989"
}

@article{Komargodski:2009rz,
    author = "Komargodski, Zohar and Seiberg, Nathan",
    title = "{From Linear SUSY to Constrained Superfields}",
    eprint = "0907.2441",
    archivePrefix = "arXiv",
    primaryClass = "hep-th",
    doi = "10.1088/1126-6708/2009/09/066",
    journal = "JHEP",
    volume = "09",
    pages = "066",
    year = "2009"
}

@article{Kuzenko:2010ef,
    author = "Kuzenko, Sergei M. and Tyler, Simon J.",
    title = "{Relating the Komargodski-Seiberg and Akulov-Volkov actions: Exact nonlinear field redefinition}",
    eprint = "1009.3298",
    archivePrefix = "arXiv",
    primaryClass = "hep-th",
    doi = "10.1016/j.physletb.2011.03.020",
    journal = "Phys. Lett. B",
    volume = "698",
    pages = "319--322",
    year = "2011"
}

@article{DallAgata:2016syy,
    author = "Dall'Agata, G. and Dudas, E. and Farakos, F.",
    title = "{On the origin of constrained superfields}",
    eprint = "1603.03416",
    archivePrefix = "arXiv",
    primaryClass = "hep-th",
    doi = "10.1007/JHEP05(2016)041",
    journal = "JHEP",
    volume = "05",
    pages = "041",
    year = "2016"
}

@article{DallAgata:2015zxp,
    author = "Dall'Agata, Gianguido and Farakos, Fotis",
    title = "{Constrained superfields in Supergravity}",
    eprint = "1512.02158",
    archivePrefix = "arXiv",
    primaryClass = "hep-th",
    reportNumber = "DFPD-2015-TH-27",
    doi = "10.1007/JHEP02(2016)101",
    journal = "JHEP",
    volume = "02",
    pages = "101",
    year = "2016"
}

@article{Khoury:2010gb,
    author = "Khoury, Justin and Lehners, Jean-Luc and Ovrut, Burt",
    title = "{Supersymmetric P(X,$\phi$) and the Ghost Condensate}",
    eprint = "1012.3748",
    archivePrefix = "arXiv",
    primaryClass = "hep-th",
    doi = "10.1103/PhysRevD.83.125031",
    journal = "Phys. Rev. D",
    volume = "83",
    pages = "125031",
    year = "2011"
}

@book{Miron:1997vjh,
    author = "Miron, Radu",
    title = "{The Geometry of Higher-Order Lagrange Spaces}",
    doi = "10.1007/978-94-017-3338-0",
    publisher = "Springer Netherlands",
    address = "Dordrecht",
    year = "1997"
}

@article{Gaiotto:2014kfa,
    author = "Gaiotto, Davide and Kapustin, Anton and Seiberg, Nathan and Willett, Brian",
    title = "{Generalized Global Symmetries}",
    eprint = "1412.5148",
    archivePrefix = "arXiv",
    primaryClass = "hep-th",
    doi = "10.1007/JHEP02(2015)172",
    journal = "JHEP",
    volume = "02",
    pages = "172",
    year = "2015"
}

@article{Wigner:1939cj,
    author = "Wigner, Eugene P.",
    editor = "Kim, Y. S. and Zachary, W. W.",
    title = "{On Unitary Representations of the Inhomogeneous Lorentz Group}",
    doi = "10.2307/1968551",
    journal = "Annals Math.",
    volume = "40",
    pages = "149--204",
    year = "1939"
}

@article{Yang:1954ek,
    author = "Yang, Chen-Ning and Mills, Robert L.",
    editor = "Hsu, Jong-Ping and Fine, D.",
    title = "{Conservation of Isotopic Spin and Isotopic Gauge Invariance}",
    doi = "10.1103/PhysRev.96.191",
    journal = "Phys. Rev.",
    volume = "96",
    pages = "191--195",
    year = "1954"
}

@article{Einstein:1916vd,
    author = "Einstein, Albert",
    editor = "Hsu, Jong-Ping and Fine, D.",
    title = "{The foundation of the general theory of relativity.}",
    doi = "10.1002/andp.19163540702",
    journal = "Annalen Phys.",
    volume = "49",
    number = "7",
    pages = "769--822",
    year = "1916"
}

@article{Nambu:1960tm,
    author = "Nambu, Yoichiro",
    editor = "Taylor, J. C.",
    title = "{Quasiparticles and Gauge Invariance in the Theory of Superconductivity}",
    doi = "10.1103/PhysRev.117.648",
    journal = "Phys. Rev.",
    volume = "117",
    pages = "648--663",
    year = "1960"
}

@article{Goldstone:1961eq,
    author = "Goldstone, J.",
    title = "{Field Theories with Superconductor Solutions}",
    doi = "10.1007/BF02812722",
    journal = "Nuovo Cim.",
    volume = "19",
    pages = "154--164",
    year = "1961"
}

@article{Weinberg:1964ew,
    author = "Weinberg, Steven",
    title = "{Photons and Gravitons in  $S$-Matrix Theory: Derivation of Charge Conservation and Equality of Gravitational and Inertial Mass}",
    doi = "10.1103/PhysRev.135.B1049",
    journal = "Phys. Rev.",
    volume = "135",
    pages = "B1049--B1056",
    year = "1964"
}

@article{Weinberg:1965nx,
    author = "Weinberg, Steven",
    title = "{Infrared photons and gravitons}",
    doi = "10.1103/PhysRev.140.B516",
    journal = "Phys. Rev.",
    volume = "140",
    pages = "B516--B524",
    year = "1965"
}

@article{Englert:1964et,
    author = "Englert, F. and Brout, R.",
    editor = "Taylor, J. C.",
    title = "{Broken Symmetry and the Mass of Gauge Vector Mesons}",
    doi = "10.1103/PhysRevLett.13.321",
    journal = "Phys. Rev. Lett.",
    volume = "13",
    pages = "321--323",
    year = "1964"
}

@article{Higgs:1964pj,
    author = "Higgs, Peter W.",
    editor = "Taylor, J. C.",
    title = "{Broken Symmetries and the Masses of Gauge Bosons}",
    doi = "10.1103/PhysRevLett.13.508",
    journal = "Phys. Rev. Lett.",
    volume = "13",
    pages = "508--509",
    year = "1964"
}

@article{Guralnik:1964eu,
    author = "Guralnik, G. S. and Hagen, C. R. and Kibble, T. W. B.",
    editor = "Taylor, J. C.",
    title = "{Global Conservation Laws and Massless Particles}",
    doi = "10.1103/PhysRevLett.13.585",
    journal = "Phys. Rev. Lett.",
    volume = "13",
    pages = "585--587",
    year = "1964"
}

@article{tHooft:1979rat,
    author = "'t Hooft, Gerard",
    editor = "'t Hooft, Gerard and Itzykson, C. and Jaffe, A. and Lehmann, H. and Mitter, P. K. and Singer, I. M. and Stora, R.",
    title = "{Naturalness, chiral symmetry, and spontaneous chiral symmetry breaking}",
    reportNumber = "PRINT-80-0083 (UTRECHT)",
    doi = "10.1007/978-1-4684-7571-5_9",
    journal = "NATO Sci. Ser. B",
    volume = "59",
    pages = "135--157",
    year = "1980"
}

@article{Seiberg:1993vc,
    author = "Seiberg, Nathan",
    title = "{Naturalness versus supersymmetric nonrenormalization theorems}",
    eprint = "hep-ph/9309335",
    archivePrefix = "arXiv",
    reportNumber = "RU-93-45",
    doi = "10.1016/0370-2693(93)91541-T",
    journal = "Phys. Lett. B",
    volume = "318",
    pages = "469--475",
    year = "1993"
}

@article{Gell-Mann:1960mvl,
    author = "Gell-Mann, Murray and Levy, M",
    title = "{The axial vector current in beta decay}",
    doi = "10.1007/BF02859738",
    journal = "Nuovo Cim.",
    volume = "16",
    pages = "705",
    year = "1960"
}

@article{Sikivie:1980hm,
    author = "Sikivie, P. and Susskind, Leonard and Voloshin, Mikhail B. and Zakharov, Valentin I.",
    title = "{Isospin Breaking in Technicolor Models}",
    reportNumber = "ITP-661-STANFORD",
    doi = "10.1016/0550-3213(80)90214-X",
    journal = "Nucl. Phys. B",
    volume = "173",
    pages = "189--207",
    year = "1980"
}

@article{Chisholm:1961tha,
    author = "Chisholm, J. S. R.",
    title = "{Change of variables in quantum field theories}",
    doi = "10.1016/0029-5582(61)90106-7",
    journal = "Nucl. Phys.",
    volume = "26",
    number = "3",
    pages = "469--479",
    year = "1961"
}

@article{Kamefuchi:1961sb,
    author = "Kamefuchi, S. and O'Raifeartaigh, L. and Salam, Abdus",
    title = "{Change of variables and equivalence theorems in quantum field theories}",
    doi = "10.1016/0029-5582(61)90056-6",
    journal = "Nucl. Phys.",
    volume = "28",
    pages = "529--549",
    year = "1961"
}

@article{Coleman:1969sm,
    author = "Coleman, Sidney R. and Wess, J. and Zumino, Bruno",
    title = "{Structure of phenomenological Lagrangians. 1.}",
    doi = "10.1103/PhysRev.177.2239",
    journal = "Phys. Rev.",
    volume = "177",
    pages = "2239--2247",
    year = "1969"
}

@article{Arzt:1993gz,
    author = "Arzt, Christopher",
    title = "{Reduced effective Lagrangians}",
    eprint = "hep-ph/9304230",
    archivePrefix = "arXiv",
    reportNumber = "UM-TH-92-28",
    doi = "10.1016/0370-2693(94)01419-D",
    journal = "Phys. Lett. B",
    volume = "342",
    pages = "189--195",
    year = "1995"
}

@article{Meetz:1969as,
    author = "Meetz, K.",
    title = "{Realization of chiral symmetry in a curved isospin space}",
    doi = "10.1063/1.1664881",
    journal = "J. Math. Phys.",
    volume = "10",
    pages = "589--593",
    year = "1969"
}

@article{Honerkamp:1971xtx,
    author = "Honerkamp, J. and Meetz, K.",
    title = "{Chiral-invariant perturbation theory}",
    doi = "10.1103/PhysRevD.3.1996",
    journal = "Phys. Rev. D",
    volume = "3",
    pages = "1996--1998",
    year = "1971"
}

@article{Honerkamp:1971sh,
    author = "Honerkamp, J.",
    title = "{Chiral multiloops}",
    doi = "10.1016/0550-3213(72)90299-4",
    journal = "Nucl. Phys. B",
    volume = "36",
    pages = "130--140",
    year = "1972"
}

@article{Ecker:1971xko,
    author = "Ecker, G. and Honerkamp, J.",
    title = "{Application of invariant renormalization to the nonlinear chiral invariant pion lagrangian in the one-loop approximation}",
    doi = "10.1016/0550-3213(71)90468-8",
    journal = "Nucl. Phys. B",
    volume = "35",
    pages = "481--492",
    year = "1971"
}

@article{Alvarez-Gaume:1981exa,
    author = "Alvarez-Gaume, Luis and Freedman, Daniel Z. and Mukhi, Sunil",
    title = "{The Background Field Method and the Ultraviolet Structure of the Supersymmetric Nonlinear Sigma Model}",
    reportNumber = "Print-81-0045 (MIT)",
    doi = "10.1016/0003-4916(81)90006-3",
    journal = "Annals Phys.",
    volume = "134",
    pages = "85",
    year = "1981"
}

@article{Alvarez-Gaume:1981exv,
    author = "Alvarez-Gaume, Luis and Freedman, Daniel Z.",
    title = "{Geometrical Structure and Ultraviolet Finiteness in the Supersymmetric Sigma Model}",
    reportNumber = "PRINT-81-0044 (MIT)",
    doi = "10.1007/BF01208280",
    journal = "Commun. Math. Phys.",
    volume = "80",
    pages = "443",
    year = "1981"
}

@article{Boulware:1981ns,
    author = "Boulware, David G. and Brown, Lowell S.",
    title = "{SYMMETRIC SPACE SCALAR FIELD THEORY}",
    reportNumber = "RLO-1388-870",
    doi = "10.1016/0003-4916(82)90192-0",
    journal = "Annals Phys.",
    volume = "138",
    pages = "392",
    year = "1982"
}

@article{Howe:1986vm,
    author = "Howe, Paul S. and Papadopoulos, G. and Stelle, K. S.",
    title = "{The Background Field Method and the Nonlinear $\sigma$ Model}",
    reportNumber = "Print-86-1354 (IAS,PRINCETON), CERN-TH-4744/87",
    doi = "10.1016/0550-3213(88)90379-3",
    journal = "Nucl. Phys. B",
    volume = "296",
    pages = "26--48",
    year = "1988"
}

@article{Dixon:1989fj,
    author = "Dixon, Lance J. and Kaplunovsky, Vadim and Louis, Jan",
    title = "{On Effective Field Theories Describing (2,2) Vacua of the Heterotic String}",
    reportNumber = "SLAC-PUB-4959, UTTG-19-89",
    doi = "10.1016/0550-3213(90)90057-K",
    journal = "Nucl. Phys. B",
    volume = "329",
    pages = "27--82",
    year = "1990"
}

@article{Alonso:2015fsp,
    author = "Alonso, Rodrigo and Jenkins, Elizabeth E. and Manohar, Aneesh V.",
    title = "{A Geometric Formulation of Higgs Effective Field Theory: Measuring the Curvature of Scalar Field Space}",
    eprint = "1511.00724",
    archivePrefix = "arXiv",
    primaryClass = "hep-ph",
    reportNumber = "CERN-PH-TH-2015-257",
    doi = "10.1016/j.physletb.2016.01.041",
    journal = "Phys. Lett. B",
    volume = "754",
    pages = "335--342",
    year = "2016"
}

@article{Alonso:2016oah,
    author = "Alonso, Rodrigo and Jenkins, Elizabeth E. and Manohar, Aneesh V.",
    title = "{Geometry of the Scalar Sector}",
    eprint = "1605.03602",
    archivePrefix = "arXiv",
    primaryClass = "hep-ph",
    reportNumber = "CERN-TH-2016-116",
    doi = "10.1007/JHEP08(2016)101",
    journal = "JHEP",
    volume = "08",
    pages = "101",
    year = "2016"
}

@article{Alonso:2016btr,
    author = "Alonso, Rodrigo and Jenkins, Elizabeth E. and Manohar, Aneesh V.",
    title = "{Sigma Models with Negative Curvature}",
    eprint = "1602.00706",
    archivePrefix = "arXiv",
    primaryClass = "hep-ph",
    reportNumber = "CERN-TH-2016-024",
    doi = "10.1016/j.physletb.2016.03.032",
    journal = "Phys. Lett. B",
    volume = "756",
    pages = "358--364",
    year = "2016"
}

@article{Helset:2018fgq,
    author = "Helset, Andreas and Paraskevas, Michael and Trott, Michael",
    title = "{Gauge fixing the Standard Model Effective Field Theory}",
    eprint = "1803.08001",
    archivePrefix = "arXiv",
    primaryClass = "hep-ph",
    doi = "10.1103/PhysRevLett.120.251801",
    journal = "Phys. Rev. Lett.",
    volume = "120",
    number = "25",
    pages = "251801",
    year = "2018"
}

@article{Nagai:2019tgi,
    author = "Nagai, Ryo and Tanabashi, Masaharu and Tsumura, Koji and Uchida, Yoshiki",
    title = "{Symmetry and geometry in a generalized Higgs effective field theory: Finiteness of oblique corrections versus perturbative unitarity}",
    eprint = "1904.07618",
    archivePrefix = "arXiv",
    primaryClass = "hep-ph",
    reportNumber = "KUNS-2755",
    doi = "10.1103/PhysRevD.100.075020",
    journal = "Phys. Rev. D",
    volume = "100",
    number = "7",
    pages = "075020",
    year = "2019"
}

@article{Finn:2019aip,
    author = "Finn, Kieran and Karamitsos, Sotirios and Pilaftsis, Apostolos",
    title = "{Frame Covariance in Quantum Gravity}",
    eprint = "1910.06661",
    archivePrefix = "arXiv",
    primaryClass = "hep-th",
    reportNumber = "MAN/HEP/2019/007",
    doi = "10.1103/PhysRevD.102.045014",
    journal = "Phys. Rev. D",
    volume = "102",
    number = "4",
    pages = "045014",
    year = "2020"
}

@article{Helset:2020yio,
    author = "Helset, Andreas and Martin, Adam and Trott, Michael",
    title = "{The Geometric Standard Model Effective Field Theory}",
    eprint = "2001.01453",
    archivePrefix = "arXiv",
    primaryClass = "hep-ph",
    doi = "10.1007/JHEP03(2020)163",
    journal = "JHEP",
    volume = "03",
    pages = "163",
    year = "2020"
}

@article{Finn:2020nvn,
    author = "Finn, Kieran and Karamitsos, Sotirios and Pilaftsis, Apostolos",
    title = "{Frame covariant formalism for fermionic theories}",
    eprint = "2006.05831",
    archivePrefix = "arXiv",
    primaryClass = "hep-th",
    reportNumber = "MAN/HEP/2020/004",
    doi = "10.1140/epjc/s10052-021-09360-w",
    journal = "Eur. Phys. J. C",
    volume = "81",
    number = "7",
    pages = "572",
    year = "2021"
}

@article{Cohen:2020xca,
    author = "Cohen, Timothy and Craig, Nathaniel and Lu, Xiaochuan and Sutherland, Dave",
    title = "{Is SMEFT Enough?}",
    eprint = "2008.08597",
    archivePrefix = "arXiv",
    primaryClass = "hep-ph",
    doi = "10.1007/JHEP03(2021)237",
    journal = "JHEP",
    volume = "03",
    pages = "237",
    year = "2021"
}

@article{Cohen:2021ucp,
    author = "Cohen, Timothy and Craig, Nathaniel and Lu, Xiaochuan and Sutherland, Dave",
    title = "{Unitarity violation and the geometry of Higgs EFTs}",
    eprint = "2108.03240",
    archivePrefix = "arXiv",
    primaryClass = "hep-ph",
    doi = "10.1007/JHEP12(2021)003",
    journal = "JHEP",
    volume = "12",
    pages = "003",
    year = "2021"
}

@article{Alonso:2021rac,
    author = "Alonso, Rodrigo and West, Mia",
    title = "{Roads to the Standard Model}",
    eprint = "2109.13290",
    archivePrefix = "arXiv",
    primaryClass = "hep-ph",
    doi = "10.1103/PhysRevD.105.096028",
    journal = "Phys. Rev. D",
    volume = "105",
    number = "9",
    pages = "096028",
    year = "2022"
}

@article{Cheung:2021yog,
    author = "Cheung, Clifford and Helset, Andreas and Parra-Martinez, Julio",
    title = "{Geometric soft theorems}",
    eprint = "2111.03045",
    archivePrefix = "arXiv",
    primaryClass = "hep-th",
    reportNumber = "CALT-TH-2021-038",
    doi = "10.1007/JHEP04(2022)011",
    journal = "JHEP",
    volume = "04",
    pages = "011",
    year = "2022"
}

@article{Cohen:2022uuw,
    author = "Cohen, Timothy and Craig, Nathaniel and Lu, Xiaochuan and Sutherland, Dave",
    title = "{On-Shell Covariance of Quantum Field Theory Amplitudes}",
    eprint = "2202.06965",
    archivePrefix = "arXiv",
    primaryClass = "hep-th",
    doi = "10.1103/PhysRevLett.130.041603",
    journal = "Phys. Rev. Lett.",
    volume = "130",
    number = "4",
    pages = "041603",
    year = "2023"
}

@article{Cheung:2022vnd,
    author = "Cheung, Clifford and Helset, Andreas and Parra-Martinez, Julio",
    title = "{Geometry-kinematics duality}",
    eprint = "2202.06972",
    archivePrefix = "arXiv",
    primaryClass = "hep-th",
    reportNumber = "CALT-TH 2022-006",
    doi = "10.1103/PhysRevD.106.045016",
    journal = "Phys. Rev. D",
    volume = "106",
    number = "4",
    pages = "045016",
    year = "2022"
}

@article{Alonso:2022ffe,
    author = "Alonso, Rodrigo and West, Mia",
    title = "{On the effective action for scalars in a general manifold to any loop order}",
    eprint = "2207.02050",
    archivePrefix = "arXiv",
    primaryClass = "hep-th",
    reportNumber = "IPPP/22/44",
    doi = "10.1016/j.physletb.2023.137937",
    journal = "Phys. Lett. B",
    volume = "841",
    pages = "137937",
    year = "2023"
}

@article{Talbert:2022unj,
    author = "Talbert, Jim",
    title = "{The geometric {\ensuremath{\nu}}SMEFT: operators and connections}",
    eprint = "2208.11139",
    archivePrefix = "arXiv",
    primaryClass = "hep-ph",
    doi = "10.1007/JHEP01(2023)069",
    journal = "JHEP",
    volume = "01",
    pages = "069",
    year = "2023"
}

@article{Helset:2022tlf,
    author = "Helset, Andreas and Jenkins, Elizabeth E. and Manohar, Aneesh V.",
    title = "{Geometry in scattering amplitudes}",
    eprint = "2210.08000",
    archivePrefix = "arXiv",
    primaryClass = "hep-ph",
    reportNumber = "CALT-TH-2022-036",
    doi = "10.1103/PhysRevD.106.116018",
    journal = "Phys. Rev. D",
    volume = "106",
    number = "11",
    pages = "116018",
    year = "2022"
}

@article{Helset:2022pde,
    author = "Helset, Andreas and Jenkins, Elizabeth E. and Manohar, Aneesh V.",
    title = "{Renormalization of the Standard Model Effective Field Theory from geometry}",
    eprint = "2212.03253",
    archivePrefix = "arXiv",
    primaryClass = "hep-ph",
    reportNumber = "CALT-TH-2022-041",
    doi = "10.1007/JHEP02(2023)063",
    journal = "JHEP",
    volume = "02",
    pages = "063",
    year = "2023"
}

@article{Craig:2023wni,
    author = "Craig, Nathaniel and Lee, Yu-Tse and Lu, Xiaochuan and Sutherland, Dave",
    title = "{Effective field theories as Lagrange spaces}",
    eprint = "2305.09722",
    archivePrefix = "arXiv",
    primaryClass = "hep-th",
    doi = "10.1007/JHEP11(2023)069",
    journal = "JHEP",
    volume = "11",
    pages = "069",
    year = "2023"
}

@article{Gattus:2023gep,
    author = "Gattus, Viola and Pilaftsis, Apostolos",
    title = "{Minimal supergeometric quantum field theories}",
    eprint = "2307.01126",
    archivePrefix = "arXiv",
    primaryClass = "hep-th",
    doi = "10.1016/j.physletb.2023.138234",
    journal = "Phys. Lett. B",
    volume = "846",
    pages = "138234",
    year = "2023"
}

@article{Assi:2023zid,
    author = "Assi, Beno{\^\i}t and Helset, Andreas and Manohar, Aneesh V. and Pag{\`e}s, Julie and Shen, Chia-Hsien",
    title = "{Fermion geometry and the renormalization of the Standard Model Effective Field Theory}",
    eprint = "2307.03187",
    archivePrefix = "arXiv",
    primaryClass = "hep-ph",
    reportNumber = "CALT-TH-2023-024, FERMILAB-PUB-23-362-T",
    doi = "10.1007/JHEP11(2023)201",
    journal = "JHEP",
    volume = "11",
    pages = "201",
    year = "2023"
}

@article{Craig:2023hhp,
    author = "Craig, Nathaniel and Lee, Yu-Tse",
    title = "{Effective Field Theories on the Jet Bundle}",
    eprint = "2307.15742",
    archivePrefix = "arXiv",
    primaryClass = "hep-th",
    doi = "10.1103/PhysRevLett.132.061602",
    journal = "Phys. Rev. Lett.",
    volume = "132",
    number = "6",
    pages = "061602",
    year = "2024"
}

@article{Alminawi:2023qtf,
    author = "Alminawi, Mohammad and Brivio, Ilaria and Davighi, Joe",
    title = "{Jet bundle geometry of scalar field theories}",
    eprint = "2308.00017",
    archivePrefix = "arXiv",
    primaryClass = "hep-ph",
    doi = "10.1088/1751-8121/ad72bb",
    journal = "J. Phys. A",
    volume = "57",
    number = "43",
    pages = "435401",
    year = "2024"
}

@article{Jenkins:2023rtg,
    author = "Jenkins, Elizabeth E. and Manohar, Aneesh V. and Naterop, Luca and Pag{\`e}s, Julie",
    title = "{An algebraic formula for two loop renormalization of scalar quantum field theory}",
    eprint = "2308.06315",
    archivePrefix = "arXiv",
    primaryClass = "hep-ph",
    reportNumber = "ZU-TH 45/23, PSI-PR-23-29",
    doi = "10.1007/JHEP12(2023)165",
    journal = "JHEP",
    volume = "12",
    pages = "165",
    year = "2023"
}

@article{Jenkins:2023bls,
    author = "Jenkins, Elizabeth E. and Manohar, Aneesh V. and Naterop, Luca and Pag{\`e}s, Julie",
    title = "{Two loop renormalization of scalar theories using a geometric approach}",
    eprint = "2310.19883",
    archivePrefix = "arXiv",
    primaryClass = "hep-ph",
    reportNumber = "ZU-TH 69/23, PSI-PR-23-39",
    doi = "10.1007/JHEP02(2024)131",
    journal = "JHEP",
    volume = "02",
    pages = "131",
    year = "2024"
}

@article{Alonso:2023jsi,
    author = "Alonso, Rodrigo and Criado, Juan Carlos and Houtz, Rachel and West, Mia",
    title = "{Walls, bubbles and doom {\textemdash} the cosmology of HEFT}",
    eprint = "2312.00881",
    archivePrefix = "arXiv",
    primaryClass = "hep-ph",
    reportNumber = "IPPP/23/74",
    doi = "10.1007/JHEP05(2024)049",
    journal = "JHEP",
    volume = "05",
    pages = "049",
    year = "2024"
}

@article{Cohen:2023ekv,
    author = "Cohen, Timothy and Lu, Xiaochuan and Sutherland, Dave",
    title = "{On amplitudes and field redefinitions}",
    eprint = "2312.06748",
    archivePrefix = "arXiv",
    primaryClass = "hep-th",
    reportNumber = "CERN-TH-2023-233",
    doi = "10.1007/JHEP06(2024)149",
    journal = "JHEP",
    volume = "06",
    pages = "149",
    year = "2024"
}

@article{Derda:2024jvo,
    author = "Derda, Maria and Helset, Andreas and Parra-Martinez, Julio",
    title = "{Soft scalars in effective field theory}",
    eprint = "2403.12142",
    archivePrefix = "arXiv",
    primaryClass = "hep-th",
    reportNumber = "CALT-TH-2024-011, CERN-TH-2024-035",
    doi = "10.1007/JHEP06(2024)133",
    journal = "JHEP",
    volume = "06",
    pages = "133",
    year = "2024"
}

@article{Helset:2024vle,
    author = "Helset, Andreas",
    title = "{Color-kinematics duality for nonlinear sigma models with nonsymmetric cosets}",
    eprint = "2406.10955",
    archivePrefix = "arXiv",
    primaryClass = "hep-th",
    reportNumber = "CERN-TH-2024-086",
    doi = "10.1103/PhysRevD.110.L101701",
    journal = "Phys. Rev. D",
    volume = "110",
    number = "10",
    pages = "L101701",
    year = "2024"
}

@article{Gattus:2024ird,
    author = "Gattus, Viola and Pilaftsis, Apostolos",
    title = "{Supergeometric quantum effective action}",
    eprint = "2406.13594",
    archivePrefix = "arXiv",
    primaryClass = "hep-th",
    doi = "10.1103/PhysRevD.110.105006",
    journal = "Phys. Rev. D",
    volume = "110",
    number = "10",
    pages = "105006",
    year = "2024"
}

@article{Cohen:2024bml,
    author = "Cohen, Timothy and Lu, Xiaochuan and Zhang, Zhengkang",
    title = "{What is the geometry of effective field theories?}",
    eprint = "2410.21378",
    archivePrefix = "arXiv",
    primaryClass = "hep-th",
    reportNumber = "CERN-TH-2024-176",
    doi = "10.1103/PhysRevD.111.085012",
    journal = "Phys. Rev. D",
    volume = "111",
    number = "8",
    pages = "085012",
    year = "2025"
}

@article{Li:2024ciy,
    author = "Li, Xu-Xiang and Lu, Xiaochuan and Zhang, Zhengkang",
    title = "{The geometric universal one-loop effective action}",
    eprint = "2411.04173",
    archivePrefix = "arXiv",
    primaryClass = "hep-ph",
    doi = "10.1007/JHEP08(2025)102",
    journal = "JHEP",
    volume = "08",
    pages = "102",
    year = "2025"
}

@article{Cohen:2024fak,
    author = "Cohen, Timothy and Forslund, Matthew and Helset, Andreas",
    title = "{Field redefinitions can be nonlocal}",
    eprint = "2412.12247",
    archivePrefix = "arXiv",
    primaryClass = "hep-th",
    reportNumber = "CERN-TH-2024-216, YITP-SB-2024-35",
    doi = "10.1007/JHEP10(2025)019",
    journal = "JHEP",
    volume = "10",
    pages = "019",
    year = "2025"
}

@article{Aigner:2025xyt,
    author = "Aigner, Patrick and Bellafronte, Luigi and Gendy, Emanuele and Haslehner, Dominik and Weiler, Andreas",
    title = "{Renormalising the field-space geometry}",
    eprint = "2503.09785",
    archivePrefix = "arXiv",
    primaryClass = "hep-th",
    doi = "10.1007/JHEP07(2025)167",
    journal = "JHEP",
    volume = "07",
    pages = "167",
    year = "2025"
}

@article{Cohen:2025dex,
    author = "Cohen, Timothy and Fadakar, Ipak and Helset, Andreas and Nardi, Filippo",
    title = "{Geometry of soft scalars at one loop}",
    eprint = "2504.12371",
    archivePrefix = "arXiv",
    primaryClass = "hep-th",
    reportNumber = "CERN-TH-2025-081",
    doi = "10.1007/JHEP08(2025)140",
    journal = "JHEP",
    volume = "08",
    pages = "140",
    year = "2025"
}

@article{Assi:2025fsm,
    author = "Assi, Beno{\^\i}t and Helset, Andreas and Pag{\`e}s, Julie and Shen, Chia-Hsien",
    title = "{Renormalizing two-fermion operators in the SMEFT via supergeometry}",
    eprint = "2504.18537",
    archivePrefix = "arXiv",
    primaryClass = "hep-ph",
    reportNumber = "CERN-TH-2025-084, FERMILAB-PUB-25-0280-V",
    doi = "10.1007/JHEP12(2025)082",
    journal = "JHEP",
    volume = "12",
    pages = "082",
    year = "2025"
}

@article{Craig:2025uoc,
    author = "Craig, Nathaniel and Lee, I-Kwan and Lee, Yu-Tse",
    title = "{Fermi geometry of the Higgs sector}",
    eprint = "2509.07101",
    archivePrefix = "arXiv",
    primaryClass = "hep-th",
    doi = "10.1007/JHEP02(2026)044",
    journal = "JHEP",
    volume = "02",
    pages = "044",
    year = "2026"
}

@article{Cohen:2025prs,
    author = "Cohen, Timothy and Li, Xu-Xiang and Zhang, Zhengkang",
    title = "{Geometric building blocks of effective field theory amplitudes}",
    eprint = "2509.20449",
    archivePrefix = "arXiv",
    primaryClass = "hep-th",
    reportNumber = "CERN-TH-2025-188",
    doi = "10.1007/JHEP02(2026)076",
    journal = "JHEP",
    volume = "02",
    pages = "076",
    year = "2026"
}

@article{Alminawi:2025pwg,
    author = "Alminawi, Mohammad and Brivio, Ilaria and Davighi, Joe",
    title = "{Scalar amplitudes from fiber bundle geometry}",
    eprint = "2509.20482",
    archivePrefix = "arXiv",
    primaryClass = "hep-th",
    reportNumber = "COMETA-2025-38, CERN-TH-2025-187",
    doi = "10.1103/qxwr-1vs6",
    journal = "Phys. Rev. D",
    volume = "113",
    number = "7",
    pages = "076005",
    year = "2026"
}

@article{Alonso:2025ksv,
    author = "Alonso, Rodrigo and Chattopadhyay, Susobhan and Ingoldby, James",
    title = "{The Potential of HEFT and the scale of New Physics}",
    eprint = "2512.13612",
    archivePrefix = "arXiv",
    primaryClass = "hep-ph",
    reportNumber = "IPPP/25/86, TIFR/TH/25-24",
    month = "12",
    year = "2025"
}

@article{Delgado:2026zpz,
    author = "Delgado, Antonio and Martin, Adam and Wang, Runqing",
    title = "{Geometric Amplitudes: A Covariant Functional Approach for Massless Scalar Theories}",
    eprint = "2604.20099",
    archivePrefix = "arXiv",
    primaryClass = "hep-th",
    month = "4",
    year = "2026"
}

@article{Kuzenko:2011ya,
    author = "Kuzenko, S. M. and McArthur, I. N.",
    title = "{Goldstino superfields for spontaneously broken N=2 supersymmetry}",
    eprint = "1105.3001",
    archivePrefix = "arXiv",
    primaryClass = "hep-th",
    doi = "10.1007/JHEP06(2011)133",
    journal = "JHEP",
    volume = "06",
    pages = "133",
    year = "2011"
}

@article{Cribiori:2016hdz,
    author = "Cribiori, N. and Dall'Agata, G. and Farakos, F.",
    title = "{Interactions of N Goldstini in Superspace}",
    eprint = "1607.01277",
    archivePrefix = "arXiv",
    primaryClass = "hep-th",
    reportNumber = "DFPD-2016-TH-10",
    doi = "10.1103/PhysRevD.94.065019",
    journal = "Phys. Rev. D",
    volume = "94",
    number = "6",
    pages = "065019",
    year = "2016"
}

@article{Dudas:2017sbi,
    author = "Dudas, E. and Ferrara, S. and Sagnotti, A.",
    title = "{A superfield constraint for $ \mathcal{N} $ = 2 {\textrightarrow} $ \mathcal{N} $ = 0 breaking}",
    eprint = "1707.03414",
    archivePrefix = "arXiv",
    primaryClass = "hep-th",
    reportNumber = "CERN-TH-2017-131, CPHT-RR043.062017",
    doi = "10.1007/JHEP08(2017)109",
    journal = "JHEP",
    volume = "08",
    pages = "109",
    year = "2017"
}

@article{Antoniadis:2010hs,
    author = "Antoniadis, I. and Dudas, E. and Ghilencea, D. M. and Tziveloglou, P.",
    title = "{Non-linear MSSM}",
    eprint = "1006.1662",
    archivePrefix = "arXiv",
    primaryClass = "hep-ph",
    reportNumber = "CERN-PH-TH-2010-118, CPHT-RR037.0510, LPT-ORSAY-10-32, NSF-KITP-10-057",
    doi = "10.1016/j.nuclphysb.2010.08.002",
    journal = "Nucl. Phys. B",
    volume = "841",
    pages = "157--177",
    year = "2010"
}

@article{Dudas:2012fa,
    author = "Dudas, Emilian and Petersson, Christoffer and Tziveloglou, Pantelis",
    title = "{Low Scale Supersymmetry Breaking and its LHC Signatures}",
    eprint = "1211.5609",
    archivePrefix = "arXiv",
    primaryClass = "hep-ph",
    reportNumber = "CERN-PH-TH-2012-317",
    doi = "10.1016/j.nuclphysb.2013.02.001",
    journal = "Nucl. Phys. B",
    volume = "870",
    pages = "353--383",
    year = "2013"
}

@article{Alonso:2014rga,
    author = "Alonso, Rodrigo and Jenkins, Elizabeth E. and Manohar, Aneesh V.",
    title = "{Holomorphy without Supersymmetry in the Standard Model Effective Field Theory}",
    eprint = "1409.0868",
    archivePrefix = "arXiv",
    primaryClass = "hep-ph",
    doi = "10.1016/j.physletb.2014.10.045",
    journal = "Phys. Lett. B",
    volume = "739",
    pages = "95--98",
    year = "2014"
}

@article{Cheung:2015aba,
    author = "Cheung, Clifford and Shen, Chia-Hsien",
    title = "{Nonrenormalization Theorems without Supersymmetry}",
    eprint = "1505.01844",
    archivePrefix = "arXiv",
    primaryClass = "hep-ph",
    reportNumber = "CALT-2015-024",
    doi = "10.1103/PhysRevLett.115.071601",
    journal = "Phys. Rev. Lett.",
    volume = "115",
    number = "7",
    pages = "071601",
    year = "2015"
}

@article{Craig:2021ksw,
    author = "Craig, Nathaniel and Garcia, Isabel Garcia and Vainshtein, Arkady and Zhang, Zhengkang",
    title = "{Magic zeroes and hidden symmetries}",
    eprint = "2112.05770",
    archivePrefix = "arXiv",
    primaryClass = "hep-ph",
    doi = "10.1007/JHEP05(2022)079",
    journal = "JHEP",
    volume = "05",
    pages = "079",
    year = "2022"
}

@article{Cohen:2016jzp,
    author = "Cohen, Timothy and Elor, Gilly and Larkoski, Andrew J.",
    title = "{Collinear Superspace}",
    eprint = "1603.09346",
    archivePrefix = "arXiv",
    primaryClass = "hep-th",
    reportNumber = "MIT-CTP-4793",
    doi = "10.1103/PhysRevD.93.125013",
    journal = "Phys. Rev. D",
    volume = "93",
    number = "12",
    pages = "125013",
    year = "2016"
}

@article{Cohen:2016dcl,
    author = "Cohen, Timothy and Elor, Gilly and Larkoski, Andrew J.",
    title = "{Soft-Collinear Supersymmetry}",
    eprint = "1609.04430",
    archivePrefix = "arXiv",
    primaryClass = "hep-th",
    reportNumber = "MIT-CTP-4826, MIT--CTP-4826",
    doi = "10.1007/JHEP03(2017)017",
    journal = "JHEP",
    volume = "03",
    pages = "017",
    year = "2017"
}

@article{Cohen:2018qvn,
    author = "Cohen, Timothy and Elor, Gilly and Larkoski, Andrew J. and Thaler, Jesse",
    title = "{Navigating Collinear Superspace}",
    eprint = "1810.11032",
    archivePrefix = "arXiv",
    primaryClass = "hep-th",
    reportNumber = "MIT--CTP 5076, MIT-CTP-5076",
    doi = "10.1007/JHEP02(2020)146",
    journal = "JHEP",
    volume = "02",
    pages = "146",
    year = "2020"
}

@article{Cohen:2019gsc,
    author = "Cohen, Timothy and Elor, Gilly and Larkoski, Andrew J. and Thaler, Jesse",
    title = "{Circumnavigating Collinear Superspace}",
    eprint = "1909.00009",
    archivePrefix = "arXiv",
    primaryClass = "hep-th",
    reportNumber = "MIT--CTP 5143",
    doi = "10.1007/JHEP02(2020)156",
    journal = "JHEP",
    volume = "02",
    pages = "156",
    year = "2020"
}

\end{document}